\documentclass[12pt]{article}
\pdfoutput=1

\usepackage{draft} 
\usepackage{hyperref}
\usepackage{graphicx,color,subfig}
\usepackage{cite}
\usepackage{mciteplus}
\usepackage{skak}
\usepackage{empheq}
\usepackage{slashed}
\usepackage{mathrsfs}
\usepackage{anyfontsize}
\usepackage{amsthm}

\usepackage{comment}
\usepackage{algorithm}
\usepackage{algpseudocode}
\usepackage{tabularx}
\usepackage{booktabs}
\usepackage{tikz}
\usetikzlibrary{arrows.meta,positioning,calc}

\DeclareFontFamily{OT1}{pzc}{}
\DeclareFontShape{OT1}{pzc}{m}{it}{<-> s * [1.10] pzcmi7t}{}
\DeclareMathAlphabet{\mathpzc}{OT1}{pzc}{m}{it}

\def\be#1\ee{\begin{align}#1\end{align}}

\newcommand{\WT}[1]{\widetilde{#1}}

\definecolor{scpurple}{RGB}{110, 90, 173}

\newcommand{\StringCode}{\textbf{string-}{\color{scpurple}\textbf{\texttt{code}}}}

\begin{document}

\unitlength = .8mm

\begin{titlepage}

\begin{center}

\hfill \\
\hfill \\
\vskip 1cm

\title{$AdS$ Leading Regge Trajectory from String Field Theory}

\author{Minjae Cho${}^a$, Jo\~{a}o Gomide${}^b$, Jaroslav Scheinpflug${}^c$, Xi Yin${}^c$}

\address{
${}^a$Leinweber Institute for Theoretical Physics, University of Chicago,\\Chicago, Illinois 60637, USA
\\
		${}^b$ICTP South American Institute for Fundamental Research
		\\
Instituto de Física Teórica, UNESP - Univ. Estadual Paulista
\\
Rua Dr. Bento Teobaldo Ferraz 271, 01140-070, São Paulo, SP, Brazil\\
${}^c$ Jefferson Physical Laboratory, Harvard University, Cambridge, MA 02138 USA
}

\email{cho7@uchicago.edu, jpr.gomide@unesp.br, jscheinpflug@g.harvard.edu, xiyin@fas.harvard.edu }

\end{center}

\abstract{We determine the $AdS$ mass of the leading Regge trajectory in superstring backgrounds $AdS_5\times S^5$, $AdS_4\times\mathbb{CP}^3$, and $AdS_3\times S^3\times{\cal M}_4$ with pure Ramond-Ramond (RR) flux for both ${\cal M}_4=S^3\times S^1$ and ${\cal M}_4=T^4$, to the second order in the expansion with respect to the background RR flux by solving the linearized equations of motion in closed superstring field theory. The $AdS_5$ and $AdS_4$ results are in a complete agreement with those obtained from the integrability approach to dual gauge theories, thereby providing a nontrivial check of the holographic duality beyond supersymmetry-protected quantities.
}

\end{titlepage}

\eject

\begingroup
\hypersetup{linkcolor=black}

\tableofcontents

\endgroup

\section{Introduction}
String theories in anti-de Sitter ($AdS$) spacetime play fundamental roles in holographic dualities \cite{Maldacena:1997re,Aharony:1999ti}. However, their first-principles formulation has been notoriously difficult since Ramond-Ramond (RR) fluxes present in most $AdS$ backgrounds made an intrinsic worldsheet formulation challenging in the standard Ramond-Neveu-Schwarz (RNS) formalism \cite{Friedan:1985ge,Friedan:1985ey,Polchinski:1998rr}.\footnote{See \cite{Berenstein:1999jq,Berenstein:1999ip} for early attempts to study RR flux backgrounds using the RNS formalism. There are alternative approaches to this problem based on Green-Schwarz and pure-spinor formalisms which may provide promising directions towards studying strings in $AdS$ \cite{Green:1983wt,Metsaev:1998it,Berkovits:2000fe,Berkovits:2004xu,Chandia:2023eel}.} Therefore, obtaining genuinely stringy physics beyond supersymmetry-protected quantities in $AdS$ backgrounds has been difficult, which made both checking the holographic duality and making predictions for the dual gauge theory practically infeasible except for special cases such as semiclassical \cite{Gubser:2002tv,Frolov:2002av,Plefka:2005bk,Roiban:2011fe} or plane-wave limits \cite{Metsaev:2001bj,Berenstein:2002jq,Metsaev:2002re}.

In recent years, closed superstring field theory (SFT) based on the RNS formalism \cite{Zwiebach:1992ie,Sen:2014dqa,Sen:2015hha,Sen:2015hia,Sen:2015uaa,deLacroix:2017lif,Sen:2024nfd} has arisen as a promising approach to studying string theories in $AdS$ spacetime \cite{Cho:2018nfn,Cho:2025coy,Alday:2026jkn}. In such a formulation, $AdS$ string backgrounds are realized as solutions to the SFT equations of motion (EOM) that manifest desired $AdS$ super-isometries. The presence of RR fluxes does not pose any challenge since they are ordinary string field components which may be turned on for background solutions. One limitation is the perturbative nature of the SFT formalism in which there are infinitely many interaction vertices in the action. Therefore, one in practice looks for $AdS$ string field solutions perturbatively order by order in the small RR-flux expansion. Despite such a limitation, the resulting $AdS$ physics described by SFT leads to the determination of genuinely stringy observables in a perturbative expansion which provides highly nontrivial checks and predictions against the dual gauge theory in the strong 't Hooft coupling regime. In short, SFT provides a perturbative \textit{definition} of strings in $AdS$.

In \cite{Cho:2025coy}, we obtained a tree-level string field solution describing type IIB strings in $AdS_5\times S^5$ background to the third order in the small RR-flux expansion around the ten-dimensional Minkowski spacetime, where the expansion parameter $\mu={1\over R}$ is identified as the inverse $AdS$-radius. We further showed that the linearized EOM around the solution in the massless sector reproduces the expected massless $AdS$ wave equation. In this work, we extend \cite{Cho:2025coy} to determine the spectrum of the leading Regge trajectory (LRT) to the second order in $\mu$. This is a first-principles string theory result which will be found to be in agreement with the integrability predictions of the dual gauge theory \cite{Gromov:2011bz,Gromov:2014bva}, providing a nontrivial check of the $AdS$/CFT correspondence. We also perform a similar analysis for other holographic backgrounds $AdS_4\times{\mathbb{CP}}^3$ and $AdS_3\times S^3\times{\cal M}_4$ with pure RR flux for ${\cal M}_4=S^3\times S^1$ and ${\cal M}_4=T^4$. For $AdS_4\times\mathbb{CP}^3$ case, LRT spectrum again agrees with the predictions from integrability \cite{Gromov:2014eha}. For $AdS_3\times S^3\times T^4$, it agrees with results from $AdS$ Virasoro-Shapiro amplitude \cite{Chester:2024wnb}, while it provides new predictions for $AdS_3\times S^3\times S^3\times S^1$.\footnote{SFT investigation of $AdS_3$ backgrounds with mixed fluxes was performed in \cite{Cho:2018nfn}, where the spectrum of pulsating string states was obtained.}

Concretely, we solve for the linearized EOM of the classical SFT that takes the form
\ie\label{eqn:linEOM}
Q\Phi+\sum_{k=1}^\infty{1\over k!}[\Psi^{\otimes k}\otimes\Phi]=0,
\fe
where string fields $\Psi$ and $\Phi$ are states of ghost number two in the worldsheet CFT for $\mathbb{R}^{1,9}$ background, $Q$ is the BRST charge, and classical string brackets $[\cdots]$ are given by flat superstring vertices of \cite{Cho:2025coy}. For the $AdS_5\times S^5$ case, $\Psi$ is the background solution obtained in \cite{Cho:2025coy}, while we will later obtain similar background solutions for the $AdS_4\times{\mathbb{CP}}^3$ and $AdS_3\times S^3\times{\cal M}_4$ cases. $\Phi$ is the linearized fluctuation encoding the string spectrum of the background $\Psi$ which we need to solve for.

We study states in the LRT labeled by the mass-level $n=1,2,\cdots$. They carry $AdS$ mass-squared $m^2_n$ and spin $s=2n+2$, but no excitations along the compact directions. The corresponding linearized fluctuation $\Phi_n$ takes the form
\ie
\Phi_n=N_n K_{a_1\cdots a_{2n+2}}(X)ce^{-\phi}\partial X^{a_1}\cdots\partial X^{a_n}\psi^{a_{n+1}} \bar c e^{-\bar\phi}\bar\partial X^{a_{n+2}}\cdots\bar\partial X^{a_{2n+1}}\bar\psi^{a_{2n+2}}+\cdots,
\fe
where $N_n={2^{n+2}\over n!}$, indices $a_1,\cdots, a_{2n+2}$ are only along $AdS$ directions, $K_{a_1\cdots a_{2n+2}}(X)$ is totally symmetric, and omitted terms include different ghost and matter oscillator structures. At the zeroth order in $\mu$, i.e. $\mathbb{R}^{1,9}$ background, BRST-closedness imposes that $K_{a_1\cdots a_{2n+2}}(X)$ is traceless and transverse in $\mathbb{R}^{1,9}$ with $m^2_n=4n$.\footnote{We use $\alpha'=1$ unit in this work.} To the second order in $\mu$, for all $AdS$ backgrounds considered in this work, we will find a tensor $T_{a_1\cdots a_{2n+2}}(X)$ that depends only on $AdS$ coordinates that is related to $K_{a_1\cdots a_{2n+2}}(X)$ by a field redefinition such that the linearized EOM \eqref{eqn:linEOM} translates to the following wave equation, transversality, and tracelessness condition in $AdS$:
\ie\label{eqn:AdSlinEOM}
[\nabla_{AdS}^2-4n-\mu^2 M_n]T_{a_1\cdots a_{2n+2}}={\cal O}(\mu^3),~~\nabla_{AdS}\cdot T_{a_1\cdots a_{2n+2}}={\cal O}(\mu^3),~~\tr_{AdS} T_{a_1\cdots a_{2n+2}}={\cal O}(\mu^3),
\fe
with
\ie\label{eqn:massCorrection}
M_n=2(n-1)(3n+1),
\fe
where $\nabla_{AdS}$ and $\tr_{AdS}$ are respectively Levi-Civita derivative and trace of the $AdS$ spacetime. Mass corrections $M_n$ correspond to corrections to the conformal dimensions of the dual gauge theory operators in the strong 't Hooft coupling expansion, which agree with the integrability \cite{Gromov:2011bz,Gromov:2014bva,Gromov:2014eha} and $AdS$ Virasoro-Shapiro amplitude \cite{Chester:2024wnb} results.

This work is organized as follows. In section \ref{sec:bgdReview}, we review the construction of the background solution $\Psi$ for $AdS_5\times S^5$ following \cite{Cho:2025coy} and similarly construct solutions for $AdS_4\times\mathbb{CP}^3$ and $AdS_3\times S^3\times {\cal M}_4$. We then investigate the linearized EOM in section \ref{sec:linEOM}, where the $AdS$ mass correction \eqref{eqn:massCorrection} is derived. Appendices collect relevant conventions and detailed computations.

\section{Review of SFT solutions for $AdS$ spacetime}\label{sec:bgdReview}
In this section, we review the $AdS$ solutions to SFT EOM derived in \cite{Cho:2025coy}. We follow worldsheet and SFT conventions introduced in \cite{Cho:2025coy}; there, see section 2 for an introduction to SFT formalism, section 3 for finding background solutions realizing desired isometries, Appendix A for worldsheet conventions, Appendix B for flat superstring vertices, and Appendix D for diffeomorphism as a string field gauge transformation. Shorter descriptions of spinor, worldsheet, and flat superstring vertex conventions are also provided in Appendices \ref{app:spinor} and \ref{app:flatV}.

Classical SFT EOM is given by
\ie\label{eqn:SFTEOMbgd}
Q\Psi+\sum_{k=2}^\infty {1\over k!}[\Psi^{\otimes k}]=0.
\fe
A solution $\Psi$ to \eqref{eqn:SFTEOMbgd} provides a classical string background. We split $\Psi$ into massless and massive components as
\ie
\Psi=\psi+(1-\mathbb{P})\Psi,~~~\psi\equiv\mathbb{P}\Psi,
\fe
where $\mathbb{P}$ projects to the $L_0^+$-nilpotent subspace. We further impose Siegel gauge condition on the massive component as
\ie
b_0^+(1-\mathbb{P})\Psi=0.
\fe
\eqref{eqn:SFTEOMbgd} then becomes the massless EOM
\ie
Q\psi+\sum_{k=2}^\infty{1\over k!}[\psi^{\otimes k}]'=0,
\fe
where $[\cdots]'$ is the massless bracket given by
\ie
[\psi^{\otimes 2}]'=\mathbb{P}[\psi^{\otimes 2}],~~~[\psi^{\otimes 3}]'=\mathbb{P}[\psi^{\otimes3}]-3\mathbb{P}\bigg[\psi\otimes h[\psi^{\otimes2}] \bigg],~~~\cdots,
\fe
with $h={b_0^+\over L_0^+}(1-\mathbb{P})$.

Super-isometries $\Lambda$ of the background solution $\Psi$ are ghost number one string fields that solve
\ie
Q\Lambda+\sum_{k=1}^\infty{1\over k!}[\Psi^{\otimes k}\otimes\Lambda]=0,
\fe
Similar to $\Psi$, we decompose $\Lambda$ into massless and massive components as
\ie
\Lambda=\lambda+(1-\mathbb{P})\Lambda,~~~\lambda\equiv\mathbb{P}\Lambda.
\fe
Then, the massless super-isometry equation becomes
\ie
Q\lambda+\sum_{k=1}^\infty{1\over k!}[\psi^{\otimes k}\otimes\lambda]'=0.
\fe

For the type IIB case, the solution $\psi$ of interest takes the form
\ie
\psi=F_{\alpha\beta}(X)V^{\alpha\beta}_{RR}+h_{\mu\nu}(X)V^{\mu\nu}_{NSNS}+\phi(X)D+A_{\mu}(X)V^\mu_{aux.},
\fe
where $F_{\alpha\beta}$ is the RR flux profile, $h_{\mu\nu}$ is the metric deformation, $\phi$ is the ghost-dilaton field, and $A_{\mu}$ is an auxiliary field. In particular, we assume that the background solution has no $B$-field. Vertex operators are defined as
\ie
{}&V_{RR}^{\alpha\beta}\equiv c\bar c e^{-{1\over2}\phi}S^{\alpha}e^{-{1\over2}\bar\phi}\bar S^\beta,~~~V_{NSNS}^{\mu\nu}\equiv4c\bar c e^{-\phi}\psi^\mu e^{-\bar\phi}\bar\psi^\nu,
\\
&D\equiv c\bar c(\eta e^{-2\bar\phi}\bar\partial\bar\xi-e^{-2\phi}\partial\xi\bar\eta),~~~V^\mu_{aux.}\equiv c^+_0(c\bar ce^{-\phi}\psi^\mu e^{-2\bar\phi}\bar\partial\bar\xi+c\bar c e^{-2\phi}\partial\xi e^{-\bar\phi}\bar\psi^\mu).
\fe
Fermionic super-isometries take the form
\ie
\lambda=\epsilon_\alpha(X)c\bar c e^{-{1\over2}\phi}S^\alpha e^{-2\bar\phi}\bar\partial\bar\xi+\bar\epsilon_\alpha (X)c\bar c e^{-2\phi}\partial\xi e^{-{1\over2}\bar\phi}\bar S^\alpha,
\fe
where $\epsilon_\alpha,\bar\epsilon_\alpha$ are Grassmann odd supersymmetry generators.

The perturbative $AdS$ solutions of interest take the form $\psi=\sum_{k=1}^\infty \mu^k\psi^{(k)}$ with
\ie
F_{\alpha\beta}=\sum_{k=1}^\infty \mu^{2k-1}F_{\alpha\beta}^{(2k-1)},~~~h_{\mu\nu}=\sum_{k=1}^\infty\mu^{2k}h_{\mu\nu}^{(2k)},~~~\phi=\sum_{k=1}^\infty\mu^{2k}\phi^{(2k)},~~~A_\mu=\sum_{k=1}^\infty\mu^{2k}A_\mu^{(2k)},
\fe
where $\mu={1\over R}$ is identified with the inverse $AdS$-radius. Then, the massless EOM to the second order in $\mu$ reads
\ie
Q\psi^{(1)}=0,~~Q\psi^{(2)}=-{1\over2}[(\psi^{(1)})^{\otimes2}]',
\fe
leading to
\ie\label{eqn:backgroundPertEOM}
{}&\Box F^{(1)}_{\alpha\beta}=0,~~(U^\mu{\partial_\mu}F^{(1)})^{\alpha}_{~\beta}=0,~~\underline{(\partial_\mu F^{(1)}U^\mu)^{~\beta}_\alpha}=0,
\\
&4\Box h^{(2)}_{\mu\nu}+\partial_\mu A^{(2)}_\nu+\partial_\nu A^{(2)}_\mu=\underline{2\Tr((F^{(1)})^TU_\mu F^{(1)}U_\nu)},
\\
&2\Box \phi^{(2)}+\partial^\mu A_\mu^{(2)}=0,~~A^{(2)}_\mu-2\partial_\mu\phi^{(2)}+4\partial^\nu h^{(2)}_{\mu\nu}=0,
\fe
where $(U_\mu)^{\alpha\beta}$ and $(L_\mu)_{\alpha\beta}$ are 16-dimensional gamma matrices defined in Appendix \ref{app:spinor}. Underlined terms will require nontrivial changes once we consider the type IIA case later.

Perturbative supersymmetries of $\Psi$ take the form $\lambda=\sum_{k=0}^\infty\mu^k\lambda^{(k)}$ leading to
\ie
\epsilon_\alpha=\sum_{k=0}^\infty\mu^k\epsilon_\alpha^{(k)},~~\bar\epsilon_\alpha=\sum_{k=0}^\infty\mu^k\bar\epsilon_\alpha^{(k)}.
\fe
We choose to focus on the case $\bar\epsilon_\alpha^{(0)}=0$ without loss of generality, since the other half with $\epsilon_\alpha^{(0)}=0$ can be constructed analogously. To the second order in $\mu$,
\ie
{}&Q\lambda^{(0)}=0,~~~Q\lambda^{(1)}=-[\psi^{(1)}\otimes\lambda^{(0)}]',
\\
&Q\lambda^{(2)}=-[\psi^{(1)}\otimes\lambda^{(1)}]'-[\psi^{(2)}\otimes\lambda^{(0)}]'-{1\over2}[(\psi^{(1)})^{\otimes2}\otimes\lambda^{(0)}]',
\fe
leading to
\ie\label{eqn:susyperteqn}
{}&\partial_\mu\epsilon_\alpha^{(0)}=0,
\\
&\Box\epsilon_\alpha^{(1)}=0,~~(U^\mu{\partial_\mu}\epsilon^{(1)})^\alpha=0,~~\partial_\mu\epsilon_\alpha^{(1)}=0,
\\
&\Box\bar\epsilon^{(1)}_\alpha=0,~~\underline{(U^\mu{\partial_\mu}\bar\epsilon^{(1)})^\alpha}=0,~~\partial_\mu\bar\epsilon_\alpha^{(1)}={i\over2}((F^{(1)})^TU_\mu\epsilon^{(0)})_\alpha,
\\
&{i\over4}(U^\mu{\partial_\mu}\epsilon^{(2)})^\alpha={3i\over16}\epsilon_\beta^{(0)}(U^\mu)^{\beta\alpha}\partial^\nu h_{\mu\nu}^{(2)}+{i\over32}{A_\mu}^{(2)}(U^\mu)^{\alpha\beta}\epsilon_\beta^{(0)},
\\
&{1\over2}\partial_\mu\epsilon^{(2)}_\alpha={1\over8}\epsilon^{(0)}_\beta(U^\nu)^{\beta\gamma}(L^\rho)_{\gamma\alpha}\partial_\rho h^{(2)}_{\mu\nu}+{3\over16}\partial_\mu\phi^{(2)}\epsilon^{(0)}_\alpha+{1\over32}\epsilon^{(0)}_\alpha A^{(2)}_\mu\underline{-{i\over 4}F^{(1)}_{\alpha\beta}(U_\mu)^{\beta\gamma}\bar\epsilon_\gamma^{(1)}},
\\
&{1\over2}\Box\epsilon^{(2)}_\alpha={3\over8}\epsilon^{(0)}_\beta(U^\nu)^{\beta\gamma}(L^\rho)_{\gamma\alpha}\partial^\mu{\partial}_\rho h^{(2)}_{\mu\nu}+{1\over16}\epsilon_\beta^{(0)}\partial_\mu{A}_\nu^{(2)}(U^\nu)^{\beta\gamma}(L^\mu)_{\gamma\alpha}.
\fe

For the type IIA case, we assume that the background RR field strength $F_\alpha^{~\beta}$ is anti-symmetric and the background $B$-field is zero, as will be the case for $AdS_4\times\mathbb{CP}^3$. With this assumption, the background and super-isometry equations for IIA can be obtained simply by making the following replacement to the IIB equations above:
\ie
F_{\alpha\beta}\rightarrow F_\alpha^{~\beta},~F^{(2k-1)}_{\alpha\beta}\rightarrow F^{(2k-1)\beta}_\alpha,~V^{\alpha\beta}_{RR}\rightarrow V^{\alpha}_{RR,\beta},~\bar S^\beta\rightarrow \bar S_\beta,~\bar\epsilon_\alpha\rightarrow\bar\epsilon^\alpha,~\bar\epsilon^{(k)}_\alpha\rightarrow\bar\epsilon^{(k)\alpha},
\fe
with appropriate index raising/lowering and the following changes for the underlined terms in \eqref{eqn:backgroundPertEOM} and \eqref{eqn:susyperteqn}
\ie
{}&\text{\eqref{eqn:backgroundPertEOM}: }~(\partial_\mu F^{(1)}U^\mu)^{~\beta}_\alpha\rightarrow (\partial_\mu F^{(1)}L^\mu)_{\alpha\beta},~~2\Tr((F^{(1)})^TU_\mu F^{(1)}U_\nu)\rightarrow-2\Tr((F^{(1)})^TU_\mu F^{(1)}L_\nu)
\\
&\text{\eqref{eqn:susyperteqn}: }~(U^\mu{\partial_\mu}\bar\epsilon^{(1)})^\alpha\rightarrow (L^\mu{\partial_\mu}\bar\epsilon^{(1)})_\alpha,~~-{i\over 4}F^{(1)}_{\alpha\beta}(U_\mu)^{\beta\gamma}\bar\epsilon_\gamma^{(1)}\rightarrow {i\over 4}F^{(1)\beta}_\alpha(L_\mu)_{\beta\gamma}(\bar\epsilon^{(1)})^\gamma.
\fe
These changes come from $\bar\psi\bar S\sim\bar S$ and $\bar S\bar S\sim\bar\psi$ OPE, where the replacement $U_\mu\rightarrow-L_\mu$ occurs as presented in Appendix \ref{app:spinor}.

\subsection{$AdS_5\times S^5$ background solution}
In \cite{Cho:2025coy}, the solution $\psi$ for $AdS_5\times S^5$ to the third order in $\mu$ was obtained. We will work only up to the second order in $\mu$ in this work, which gives
\ie\label{eqn:AdS5bgdSol}
{}&F_{\alpha\beta}^{(1)}(X)=i(L^0U^1L^2U^3L^4)_{\alpha\beta},~~~h_{\mu\nu}^{(2)}(X)=-[\delta^a_\mu\delta^b_\nu (X_a X_b-4\kappa\eta_{ab})-\delta^i_\mu\delta^j_\nu (X_iX_j-4\kappa\delta_{ij})],
\\
&\phi^{(2)}(X)=-X^2,~~~A_\mu^{(2)}(X)=10\partial_\mu X^2,~~~~~~X^2=X_aX^a-X_iX^i,
\fe
where $\mu,\nu=0,\cdots,9$ are $\mathbb{R}^{1,9}$ indices that we divide into $a,b=0,\cdots,4$ for $AdS_5$ and $i,j=5,\cdots,9$ for $S^5$. Here, $\kappa=\tau+{b\over4}$ with $\tau=\log r_0$ and $b=\log(2z_0)$, where $r_0$ and $z_0$ are SFT parameters that appear in the flat superstring vertex choice (see (\ref{eqn:flatCoordMapr0z0}) and (\ref{eqn:flatCoordMapr0z0p}) in Appendix \ref{app:flatV}). $\kappa$-dependent terms in $h^{(2)}_{\mu\nu}$ will make later manipulations of linearized EOM more convenient by canceling against local chart effects and $F^{(1)}_{\alpha\beta}(X)$ contributions. Note that they are $Q$-exact:\footnote{In \cite{Cho:2025coy}, the presence of the $\tau$ term was attributed to Hata-Zwiebach double bracket \cite{Hata:1993gf} (see \eqref{eqn:hataZwiebach}). In contrast, $b$-dependent term was absent. In fact, string bracket computation (F.2) in \cite{Cho:2025coy} missed the dependence on $z_0$ originating from the contraction of $X$ zero-modes. Once this is correctly included, massless wave equation for the axion can be obtained either by including the $b$-dependent term in $h^{(2)}_{\mu\nu}$ of \eqref{eqn:AdS5bgdSol} or applying further field redefinitions to the axion field beyond the ones already performed in \cite{Cho:2025coy}.}
\ie
{}&Q \cdot c\bar c\bigg[e^{-\phi}(X_a\psi^a-X_i\psi^i)e^{-2\bar\phi}\bar\partial\bar\xi+e^{-2\phi}\partial\xi e^{-\bar\phi}(X_a\bar\psi^a-X_i\bar\psi^i) \bigg]\\
&=c\bar c\left(\eta_{ab} e^{-\phi}\psi^a e^{-\bar\phi}\bar\psi^b-\delta_{ij} e^{-\phi}\psi^i e^{-\bar\phi}\bar\psi^j \right).
\fe
Apart from the $\kappa$-dependent part, the metric profile is nothing but the $AdS_5\times S^5$ metric in the embedding coordinates expanded to the second order in $\mu$ (see Appendix \ref{app:AdSsugra}) and the physical dilaton ${1\over4}(h_\mu^\mu-\phi)$ vanishes to this order as desired. The bosonic sub-isometry $SO(1,4)\times SO(5)$ is manifest and \cite{Cho:2025coy} further showed that this solution preserves 32 supersymmetries. Indeed, the solution to \eqref{eqn:susyperteqn} is given by
\ie
\epsilon_\alpha^{(0)}=\text{const.},~~\epsilon_\alpha^{(1)}=0,~~\bar\epsilon^{(1)}_\alpha={i\over2}X^\mu (F^{(1)})^T_{\alpha\beta}U_\mu^{\beta\gamma}\epsilon^{(0)}_\gamma,
\fe
and equations for $\epsilon^{(2)}_\alpha$ were shown to be integrable for all components $\alpha$ given the background solution \eqref{eqn:AdS5bgdSol}.

\subsection{$AdS_4\times\mathbb{CP}^3$ background solution}
It is straightforward to extend the construction to IIA $AdS_4\times\mathbb{CP}^3$ case. Using the embedding coordinates for $AdS_4$ and coordinates for $\mathbb{CP}^3$ presented in Appendix \ref{app:AdSsugra}, the background solution to the second order in $\mu$ is given by
\ie\label{eqn:AdS4bgdsol}
{}&F^{(1)\beta}_\alpha=-{i\over4}\left(L^4U^5+L^6U^7+L^8U^9+3L^0U^1L^2U^3 \right)_\alpha^{~\beta},
\\
&h^{(2)}_{\mu\nu}(X)=-\delta^a_\mu\delta^b_\nu[X_aX_b-(3\tau+b)\eta_{ab}]+\delta_\mu^i\delta_\nu^j\left[{1\over4}(X_iX_j-(JX)_i(JX)_j)-2\tau\delta_{ij} \right],
\\
&\phi^{(2)}(X)=-X_aX^a+4b,~~A^{(2)}_\mu=\delta_\mu^a16X_a-\delta_\mu^i8X_i.
\fe
As before, we introduced SFT parameter dependent shifts for later convenience which is $Q$-exact:
\ie
{}&Q\left[4c\bar c\left((3\tau+b)X_a(e^{-\phi}\psi^a e^{-2\bar\phi}\bar\partial\bar\xi+e^{-2\phi}\partial\xi e^{-\bar\phi}\bar\psi^a)-2\tau X_i(e^{-\phi}\psi^i e^{-2\bar\phi}\bar\partial\bar\xi+e^{-2\phi}\partial\xi e^{-\bar\phi}\bar\psi^i) \right) \right]
\\
&=4c\bar c\left[ (3\tau+b)e^{-\phi}\psi^a e^{-\bar\phi}\bar\psi_a-2\tau e^{-\phi}\psi^i e^{-\bar\phi}\bar\psi_i+b(\eta e^{-2\bar\phi}\bar\partial\bar\xi-e^{-2\phi}\partial\xi\bar\eta) \right].
\fe
Supersymmetry generators (with $\bar\epsilon^{(0)\alpha}=0$) are
\ie
{}&\epsilon^{(0)}_\alpha=P_{12}\epsilon^{(0)}_\alpha=\text{const.},~~\epsilon^{(1)}_\alpha=0,~~\bar\epsilon^{(1)\alpha}={i\over2}X^\mu\left((F^{(1)})^T\right)^\alpha_{~\beta}U_\mu^{\beta\gamma}\epsilon^{(0)}_\gamma
\\
&\epsilon^{(2)}_\alpha=-{X_\mu X^\nu\over8}F^{(1)\beta}_\alpha L^\mu_{\beta\gamma}\left((F^{(1)})^T\right)^{\gamma}_{~\delta}U_\nu^{\delta\lambda} \epsilon^{(0)}_\lambda-{1\over2}X_aX^a\epsilon^{(0)}_\alpha,~~\bar\epsilon^{(2)}=0.
\fe
Here, $P_{12}$ is a projector to a 12-dimensional subspace of 16-dimensional constant spinors for $\epsilon^{(0)}_\alpha$. This constraint comes from $(L^\mu{\partial_\mu}\bar\epsilon^{(1)})_\alpha=0$ that requires
\ie\label{eqn:24susy}
L^\mu_{\alpha\beta}\left((F^{(1)})^T\right)^\beta_{~\gamma}U_\mu^{\gamma\delta}\epsilon^{(0)}_\delta=0.
\fe
One can immediately show that the nullity of the matrix $L^\mu F^{(1)} U_\mu$ is 12, leading to 12 supersymmetry generators. Analogous construction with $\epsilon^{(0)}_\alpha=0$ provides the other 12 generators, showing that the solution preserves 24 supersymmetries to this order as desired.

\subsection{$AdS_3\times S^3\times{\cal M}_4$ background solution}
We consider $AdS_3\times S^3\times{\cal M}_4$ for both ${\cal M}_4= S^3\times S^1$ and ${\cal M}_4= T^4$ backgrounds with pure RR flux. See Appendix \ref{app:AdSsugra} for the supergravity backgrounds. The worldsheet CFT can be taken to be $U(1)$ along the 9-th direction for the former and $U(1)^4$ along the $6,7,8,9$-th directions for the latter. We begin with ${\cal M}_4=S^3\times S^1$, whose background solution to the second order in $\mu$ is given by
\ie\label{eqn:AdS3bgdsol}
{}&F_{\alpha\beta}^{(1)}(X)={i\over2}(L^0U^1L^2+p_+L^3U^4L^5+p_-L^6U^7L^8)_{\alpha\beta},~~~
\\
&h_{\mu\nu}^{(2)}(X)=-[\delta^a_\mu\delta^b_\nu (X_a X_b-\kappa_3\eta_{ab})-\delta^i_\mu\delta^j_\nu p_+^2(X_iX_j-\kappa_3\delta_{ij})-\delta^r_\mu\delta^s_\nu p_-^2(X_rX_s-\kappa_3\delta_{rs})],
\\
&\phi^{(2)}(X)=-X_aX^a+p_+^2X_iX^i+p_-^2X_rX^r,~~~A_\mu^{(2)}(X)=12\left[\delta_\mu^a X_a-\delta_\mu^ip_+^2X_i-\delta_\mu^rp_-^2X_r \right],
\fe
where $\mu,\nu$ indices are divided into $a,b=0,1,2$ for $AdS_3$, $i,j=3,4,5$ for $S^3$, $r,s=6,7,8$ for another $S^3$, and the last index for $S^1$ which we do not display. $\kappa_3=2\tau+b$ and $p_\pm^2={R^2\over R_\pm^2}$ with $p_+^2+p_-^2=1$, where $R_\pm$ is the radii of two $S^3$'s. $\kappa_3$-dependent terms are $Q$-exact:
\ie
{}&Q \cdot c\bar c\bigg[e^{-\phi}(X_a\psi^a-p_+^2X_i\psi^i-p_-^2X_r\psi^r)e^{-2\bar\phi}\bar\partial\bar\xi+e^{-2\phi}\partial\xi e^{-\bar\phi}(X_a\bar\psi^a-p_+^2X_i\bar\psi^i-p_-^2X_r\bar\psi^r) \bigg]\\
&=c\bar c\left(\eta_{ab} e^{-\phi}\psi^a e^{-\bar\phi}\bar\psi^b-p_+^2\delta_{ij} e^{-\phi}\psi^i e^{-\bar\phi}\bar\psi^j-p_-^2\delta_{rs} e^{-\phi}\psi^r e^{-\bar\phi}\bar\psi^s \right).
\fe
Supersymmetry generators with $\bar\epsilon_\alpha^{(0)}=0$ are
\ie
\epsilon^{(0)}_\alpha=&P_{8}\epsilon^{(0)}_\alpha=\text{const.},~~\epsilon^{(1)}_\alpha=0,~~\bar\epsilon^{(1)}_\alpha={i\over2}X^\mu\left((F^{(1)})^T\right)_{\alpha\beta}U_\mu^{\beta\gamma}\epsilon^{(0)}_\gamma,
\\
\epsilon^{(2)}_\alpha=&{3\over8}(-X_aX^a+p_+^2X_iX^i+p_-^2X_rX^r)\epsilon^{(0)}_\alpha+{1\over4}\bigg[X^aX^i\left(F^{(1)}U_a (F^{(1)})^TU_i\right)_\alpha^{~\beta}
\\
&+X^aX^r\left(F^{(1)}U_a (F^{(1)})^TU_r\right)_\alpha^{~\beta}+X^iX^r\left(F^{(1)}U_i (F^{(1)})^TU_r\right)_\alpha^{~\beta}  \bigg]\epsilon^{(0)}_\beta,
\\
\bar\epsilon^{(2)}=&0,
\fe
where the projection to the 8-dimensional constant spinor space $P_8$ originates from
\ie
\left(U^\mu (F^{(1)})^T U_\mu \right)^{\alpha\beta}\epsilon_\beta^{(0)}=0,
\fe
with the nullity of the matrix $U^\mu (F^{(1)})^T U_\mu$ being 8. Analogous solutions with $\epsilon^{(0)}_\alpha=0$ gives the other 8 generators, leading to in total of 16 supersymmetries as expected.

The case of ${\cal M}_4=T^4$ can be simply obtained by setting $p_-=0$ in the above and replacing the 6,7,8,9-th directions by $U(1)^4$ worldsheet CFT. Since observables of interest in this work will not be sensitive to these internal directions, we will focus on ${\cal M}_4=S^3\times S^1$ for the rest of this work and specialize to $p_-=0$ to obtain results for ${\cal M}_4=T^4$ case.

\section{Spectrum of leading Regge trajectory}\label{sec:linEOM}
In this section, we present the computation of the LRT spectrum for $AdS$ backgrounds reviewed in section \ref{sec:bgdReview}. Given such background solutions $\Psi$, the string spectrum is obtained by solving for the linearized EOM \eqref{eqn:linEOM}. We decompose the linearized fluctuation $\Phi$ into mass-level $n$ part and the rest as
\ie
\Phi=\varphi_n+(1-\mathbb{P}_n)\Phi,~~~\varphi_n=\mathbb{P}_n\Phi,
\fe
where $\mathbb P_n$ projects to the space of states with oscillator level $n$ on both holomorphic and anti-holomorphic sides multiplied by functions of $X$ zero-modes. We impose the Siegel gauge condition on the components projected out by $\mathbb P_n$ as $b_0^+(1-{\mathbb P_n})\Phi=0$. The linearized EOM then becomes
\ie
Q\varphi_n+\sum_{k=1}^\infty{1\over k!}[\psi^{\otimes k}\otimes\varphi_n]_n'=0,
\fe
where $[\cdots]_n'$ is the mass-level $n$ effective bracket given by
\ie
{}&[\psi\otimes V]_n'=\mathbb{P}_n[\psi\otimes V],
\\
&[\psi^{\otimes2}\otimes V]_n'=\mathbb{P}_n\left( [\psi^{\otimes2}\otimes V]-[h[\psi^{\otimes2}]\otimes V]-2[\psi\otimes h_n[\psi\otimes V]] \right),
\fe
for a mass-level $n$ string field $V$, with $h_n={b_0^+\over L_0^+}(1-\mathbb{P}_n)$. Expanding $\varphi_n$ in $\mu$ as
\ie
\varphi_n=\sum_{k=0}^\infty\mu^k\varphi^{(k)}_n,
\fe
the linearized EOM to the second order in $\mu$ is written as
\ie\label{eqn:linEOMpert}
{}&Q\varphi_n^{(0)}=0,~~~Q\varphi^{(1)}_n+[\psi^{(1)}\otimes\varphi^{(0)}_n]_n'=0,
\\
&Q\varphi^{(2)}_n+[\psi^{(2)}\otimes\varphi^{(0)}_n]_n'+[\psi^{(1)}\otimes\varphi^{(1)}_n]_n'+{1\over2}[(\psi^{(1)})^{\otimes2}\otimes\varphi^{(0)}_n]_n'=0.
\fe

\subsection{Leading Regge trajectory: zeroth order in $\mu$}
We first consider states in LRT at the zeroth order in $\mu$, i.e. $\mathbb{R}^{1,9}$. They take the form
\ie\label{eqn:LRT0th}
\varphi^{(0)}_n=N_n T^{(0)}_{a_1\cdots a_{2n+2}}(X)ce^{-\phi}\partial X^{a_1}\cdots\partial X^{a_n}\psi^{a_{n+1}} \bar c e^{-\bar\phi}\bar\partial X^{a_{n+2}}\cdots\bar\partial X^{a_{2n+1}}\bar\psi^{a_{2n+2}},
\fe
where $N_n={2^{n+2}\over n!}$. We are going to consider the case where the indices $a_l$ are only along the $AdS$ directions and $T^{(0)}_{a_1\cdots a_{2n+2}}(X)$ depends only on the $AdS$ coordinates. $T^{(0)}_{a_1\cdots a_{2n+2}}(X)$ is further taken to be totally symmetric, leading to the maximal spacetime spin $s=2n+2$ along the $AdS$ directions. The first equation of \eqref{eqn:linEOMpert} gives
\ie
Q\varphi^{(0)}_n={\cal J}_w[(\Box-4n)T^{(0)}]+{\cal J}_d[\partial\cdot T^{(0)}]+{\cal J}_t[\tr_\eta T^{(0)}]=0,
\fe
where ${\cal J}$'s are maps from totally symmetric tensors to ghost-number 3 string fields and $\tr_\eta$ is trace-contraction by $\eta^{ab}$. For totally symmetric tensors $\cal U$, $\cal D$, and $\cal A$ of ranks $2n+2,2n+1$, and $2n$ respectively, ${\cal J}$'s are defined as
\ie
{\cal J}_w[{\cal U}]=&-{N_n\over2}c_0^+  {\cal U}_{a_1\cdots a_{2n+2}}(X) ce^{-\phi}\partial X^{a_1}\cdots\partial X^{a_n}\psi^{a_{n+1}} \bar c e^{-\bar\phi}\bar\partial X^{a_{n+2}}\cdots\bar\partial X^{a_{2n+1}}\bar\psi^{a_{2n+2}},
\\
{\cal J}_d[{\cal D}]=&{N_n\over4}{\cal D}_{a_1\cdots a_n b_1\cdots b_{n+1}}(X)\bigg( \left[nc\partial^2ce^{-\phi}\psi^{a_1}\prod_{r=2}^n\partial X^{a_r}-c\eta\prod_{r=1}^n\partial X^{a_r} \right]\bar B_n^{b_1\cdots b_{n+1}}
\\
&-B_n^{b_1\cdots b_{n+1}}\left[ n\bar c\bar\partial^2\bar ce^{-\bar\phi}\bar\psi^{a_1}\prod_{r=2}^n\bar\partial X^{a_r}-\bar c\bar\eta\prod_{r=1}^n\bar\partial X^{a_r} \right]\bigg),
\\
{\cal J}_t[{\cal A}]=&N_n{\cal A}_{a_1\cdots a_{n-1}b_1\cdots b_{n+1}}(X)\bigg(\left[ {n(n-1)\over24}c\partial^3 ce^{-\phi}\psi^{a_1}\prod_{r=2}^{n-1}\partial X^{a_r}-{n\over4}(c\partial\eta+c\eta\partial\phi)\prod_{r=1}^{n-1}\partial X^{a_r} \right]\bar B_n^{b_1\cdots b_{n+1}}
\\
&-B_n^{b_1\cdots b_{n+1}}\left[ {n(n-1)\over24}\bar c\bar\partial^3\bar ce^{-\bar\phi}\bar\psi^{a_1}\prod_{r=2}^{n-1}\bar\partial X^{a_r}-{n\over4}(\bar c\bar\partial\bar\eta+\bar c\bar\eta\bar\partial\bar\phi)\prod_{r=1}^{n-1}\bar\partial X^{a_r} \right]\bigg),
\fe
where $\prod_{r=p}^q(\cdots)=1$ if $q<p$ (which happens for $n=1,2$) and
\ie
B_n^{b_1\cdots b_{n+1}}=c e^{-\phi}\psi^{b_1}\prod_{r=2}^{n+1}\partial X^{b_r}.
\fe

We arrive at
\ie\label{eqn:0thEqnSol}
(\Box-4n)T^{(0)}_{a_1\cdots a_{2n+2}}=0,~~~\partial^{a_1}T^{(0)}_{a_1\cdots a_{2n+2}}=0,~~~\eta^{a_1a_2}T^{(0)}_{a_1\cdots a_{2n+2}}=0,
\fe
respectively the wave equation, transversality, and tracelessness. Henceforth, we work with the following solution that carries spin $s=2n+2$ along the 1-2 plane:
\ie\label{eqn:LRT0thsol}
T^{(0)}_{a_1\cdots a_{2n+2}}(X)=u_{a_1}\cdots u_{a_{2n+2}}e^{ik\cdot X},~~~k^\mu=(E,0,\cdots,0),~~~u^\mu=(0,1,i,0,\cdots,0),
\fe
with $E=2\sqrt{n}$.

\subsection{Leading Regge trajectory: first order in $\mu$}
At the first order, the string bracket that enters the linearized EOM is
\ie\label{eqn:LRTfirstorder}
[\psi^{(1)}\otimes\varphi^{(0)}_n]_n'.
\fe
In all of $AdS$ backgrounds under consideration, $\psi^{(1)}$ lies purely in the RR sector and carries zero spin along the $AdS$ directions. Therefore, the bracket \eqref{eqn:LRTfirstorder} leads to a string field carrying maximal $AdS$ spin $2n+2$ along 1-2 plane at mass-level $n$ while being in the RR sector. However, there is no such state since the maximal spin of a mass-level $n$ state in RR sector is $2n+1$ due to the presence of spin fields carrying spins $\pm{1\over2}$. We thus conclude that $[\psi^{(1)}\otimes\varphi^{(0)}_n]_n'=0$, and we may choose any $\varphi_n^{(1)}$ such that
\ie
Q\varphi_n^{(1)}=0.
\fe

For later comparison against the $AdS$ wavefunction in section \ref{sec:AdSwave}, we choose the solution
\ie\label{eqn:LRT1stsol}
\varphi^{(1)}_n=N_n T^{(1)}_{a_1\cdots a_{2n+2}}(X)ce^{-\phi}\partial X^{a_1}\cdots\partial X^{a_n}\psi^{a_{n+1}} \bar c e^{-\bar\phi}\bar\partial X^{a_{n+2}}\cdots\bar\partial X^{a_{2n+1}}\bar\psi^{a_{2n+2}},
\fe
with
\ie\label{eqn:LRT1stTsol}
{}&T^{(1)}_{a_1\cdots a_{2n+2}}=e^{-iEt}\left[\left(-{E\rho^2\over2}-{i(d-1)t\over2}\right)u_{a_1}\cdots u_{a_{2n+2}}+{is\over E}zk_{(a_1}u_{a_2}\cdots u_{a_{2n+2})} \right],
\\
&t=X^0,~~z=X^1+i X^2,~~\rho^2=\sum_{a=1}^{d-1}(X^a)^2,
\fe
where $d=3,4,5$ is the dimension of the $AdS$ in our solutions \eqref{eqn:AdS5bgdSol} and \eqref{eqn:AdS4bgdsol}. This satisfies
\ie\label{eqn:1stOrderEqSol}
(\Box-4n)T^{(1)}_{a_1\cdots a_{2n+2}}=0,~~~\partial^{a_1}T^{(1)}_{a_1\cdots a_{2n+2}}=0,~~~\eta^{a_1a_2}T^{(1)}_{a_1\cdots a_{2n+2}}=0.
\fe
In fact, $T^{(1)}$ is related to $T^{(0)}$ in \eqref{eqn:LRT0thsol} as
\ie
T^{(1)}={E\over2}\sum_{a=1}^{d-1}{\partial^2 T^{(0)}(\textbf{p})\over\partial \textbf{p}_a^2}\bigg|_{\textbf{p}=0}-{s\over2E}T^{(0)},
\fe
where $T^{(0)}(\textbf{p})$ is $T^{(0)}$ in a boosted frame where energy and spatial momenta of $AdS$ are given respectively by $\sqrt{E^2+\textbf{p}^2}$ and $\textbf{p}$.\footnote{Concretely, $u_0(\textbf{p})=-{p_1+ip_2\over E},~u_a(\textbf{p})=\delta_{a1}+i\delta_{a2}+{p_a(p_1+ip_2)\over E(\sqrt{E^2+\textbf{p}^2}+E)},~~a=1,\cdots,d-1$.} This leads to the vanishing of the bracket
\ie
[\psi^{(1)}\otimes\varphi^{(1)}_n]_n'=0,
\fe
using the same argument as the vanishing of $[\psi^{(1)}\otimes\varphi^{(0)}_n]_n'=0$ above. Therefore, this term does not contribute to the order $\mu^2$ part of the linearized EOM \eqref{eqn:linEOMpert} that we discuss now.

\subsection{Leading Regge trajectory: second order in $\mu$}
With $[\psi^{(1)}\otimes\varphi^{(0)}_n]_n'=0$, we arrive at the second order linearized EOM given by
\ie\label{eqn:2ndLinEOM}
Q\varphi^{(2)}_n+S^{NS}_n+S^{R}_n=0,~~~S_n^{NS}=[\psi^{(2)}\otimes\varphi^{(0)}_n]_n',~~~S_n^{R}={1\over2}[(\psi^{(1)})^{\otimes2}\otimes\varphi^{(0)}_n]_n'.
\fe
Our goal is to rewrite \eqref{eqn:2ndLinEOM} in the form of \eqref{eqn:AdSlinEOM} to obtain the $AdS$ mass correction to the second order in $\mu$. We begin by performing a couple of field redefinitions that reorganizes \eqref{eqn:2ndLinEOM}.

First, we define rescaled string fields for the linearized fluctuation $\varphi_n^{(2)}$ and source terms $S_n^Y$ for $Y=NS,R$:
\ie\label{eqn:rescaling2nd}
\hat\varphi_n^{(2)}=e^{-A(\Box-4n)}\varphi^{(2)}_n,~~~\hat S_n^Y=e^{-A(\Box-4n)}S_n^{Y},~~~Y=NS,R,
\fe
where $A={\tau-b\over2}$ and $\Box$ is the Laplacian in $\mathbb{R}^{1,9}$. The term $e^{-A(\Box-4n)}$ commutes with $Q$ so that \eqref{eqn:2ndLinEOM} becomes
\ie
Q\hat\varphi_n^{(2)}+\hat S_n^{NS}+\hat S_n^{R}=0.
\fe
This rescaling removes one universal local coordinate chart effect associated with the `outgoing' string field which results from flat superstring vertex in $S_n^{NS}$. To see this, consider working in the momentum space such that the background string field profile $f(X)$ of $\psi^{(2)}$, such as $h^{(2)}_{\mu\nu},\phi^{(2)}(X),A_\mu^{(2)}(X)$, is written as $f(x)=\int {d^{10}p\over(2\pi)^{10}}\tilde f(p)e^{ip\cdot X}$. The overlap of $S_n^{NS}$ with a state $V_q$ at mass-level $n$ carrying momentum $q$ given by $\langle V_q|c_0^-|S_n^{NS}\rangle$ has a $X$ zero-mode contraction which gives an overall factor $r_0^{-(q^2+4n)/2}(2 z_0)^{k\cdot p}=e^{-A(q^2+4n)}e^{-bp^2/2}$, where $k$ is the momentum of $\varphi^{(0)}_n$ satisfying $k^2=-4n$ and we used $q=-p-k$. Back in the position space, this translates to $e^{A(\Box_{out}-4n)}e^{b\Box_{bg}/2}$, where $\Box_{out}$ and $\Box_{bg}$ are respectively $\mathbb{R}^{1,9}$ Laplacian acting on $V_q$ and background field profile in $\psi^{(2)}$. Therefore, the rescaling \eqref{eqn:rescaling2nd} removes the overall factor $e^{A(\Box_{out}-4n)}$. In contrast, $e^{b\Box_{bg}/2}$ term still remains and its effect is to cancel $b$-dependent terms in the solution $\psi^{(2)}$ in \eqref{eqn:AdS5bgdSol} and \eqref{eqn:AdS4bgdsol}. Explicitly,
\ie\label{eqn:bDependence}
AdS_5\times S^5:~~&e^{b\Box_{bg}/2}h^{(2)}_{\mu\nu}(X)=-[\delta^a_\mu\delta^b_\nu (X_a X_b-4\tau\eta_{ab})-\delta^i_\mu\delta^j_\nu (X_iX_j-4\tau\delta_{ij})],
\\
&e^{b\Box_{bg}/2}\phi^{(2)}(X)=\phi^{(2)}(X),~~e^{b\Box_{bg}/2}A_\mu^{(2)}(X)=A_\mu^{(2)}(X),
\\
AdS_4\times\mathbb{CP}^3:~~&e^{b\Box_{bg}/2}h^{(2)}_{\mu\nu}(X)=-\delta^a_\mu\delta^b_\nu[X_aX_b-3\tau\eta_{ab}]+\delta_\mu^i\delta_\nu^j\left[{1\over4}(X_iX_j-(JX)_i(JX)_j)-2\tau\delta_{ij} \right],
\\
&e^{b\Box_{bg}/2}\phi^{(2)}(X)=-X_aX^a,~~e^{b\Box_{bg}/2}A_\mu^{(2)}(X)=A_\mu^{(2)}(X),
\\
AdS_3\times S^3\times {\cal M}_4:~~&e^{b\Box_{bg}/2}h^{(2)}_{\mu\nu}(X)=-[\delta^a_\mu\delta^b_\nu (X_a X_b-2\tau\eta_{ab})-\delta^i_\mu\delta^j_\nu p_+^2(X_iX_j-2\tau\delta_{ij})
\\
&~~~~~~~~~~~~~~~~~~~~~~~-\delta^r_\mu\delta^s_\nu p_-^2(X_rX_s-2\tau\delta_{rs})],
\\
&e^{b\Box_{bg}/2}\phi^{(2)}(X)=\phi^{(2)}(X),~~e^{b\Box_{bg}/2}A_\mu^{(2)}(X)=A_\mu^{(2)}(X),
\fe
where ${\cal M}_4=S^3\times S^1$ and the resulting expressions are independent of $b$.

The second field redefinition is motivated by the diffeomorphism analysis of Appendix \ref{app:Diffeo}. Define ${\cal V}_n$ that maps a totally symmetric rank-$(2n+2)$ tensor ${\cal U}_{a_1\cdots a_{2n+2}}(X)$ to a string field in $\mathbb{R}^{1,9}$ as
\ie
{\cal V}_n[{\cal U}]=N_n {\cal U}_{a_1\cdots a_{2n+2}}(X) ce^{-\phi}\partial X^{a_1}\cdots\partial X^{a_n}\psi^{a_{n+1}} \bar c e^{-\bar\phi}\bar\partial X^{a_{n+2}}\cdots\bar\partial X^{a_{2n+1}}\bar\psi^{a_{2n+2}}.
\fe
In SFT, coefficient fields like ${\cal U}$ do not transform covariantly under SFT gauge transformations that are interpreted as diffeomorphisms. As discussed in \cite{Mazel:2025fxj}, the coefficient $h_{\mu\nu}(X)$ of $V^{\mu\nu}_{NSNS}$ for example is nonlinearly related to the metric field which transforms covariantly. With the goal of translating the linearized EOM into the covariant wave equation, transversality, and tracelessness in $AdS$ as in \eqref{eqn:AdSlinEOM}, we introduce a field redefinition of the form
\ie
{\cal U}_{a_1\cdots a_{2n+2}}(X)=T_{a_1\cdots a_{2n+2}}(X)+\left(\mathcal{F}_h(T(X))\right)_{a_1\cdots a_{2n+2}}+{\cal O}(h^2),
\fe
where $T_{a_1\cdots a_{2n+2}}(X)$ transforms covariantly under diffeomorphism and $\mathcal{F}_h(T)$ is a field redefinition linear in $h_{\mu\nu}$. For the second order linearized EOM at hand, we write
\ie\label{eqn:VClinFluc}
\hat\varphi^{(2)}_n={\cal V}_n\left[T^{(2)}+F_H(T^{(0)})\right]+{\cal C}_n(T^{(0)}),
\fe
with
\ie\label{eqn:Texpansion}
T_{a_1\cdots a_{2n+2}}(X)=T^{(0)}_{a_1\cdots a_{2n+2}}(X)+\mu T^{(1)}_{a_1\cdots a_{2n+2}}(X)+\mu^2 T^{(2)}_{a_1\cdots a_{2n+2}}(X)+{\cal O}(\mu^3),
\fe
where $F_H(T^{(0)})$ is the field redefinition term related to ${\cal F}_h(T)$ by setting $h_{\mu\nu}=\mu^2 H_{\mu\nu}$ and is linear in the product of $H_{\mu\nu}$ and $T^{(0)}$\footnote{Here, $H_{\mu\nu}$ for our specific $AdS$ backgrounds will be the SFT parameter independent part of the second order metric profile $h_{\mu\nu}^{(2)}$, i.e. $H_{\mu\nu}=-[\delta^a_\mu\delta^b_\nu X_aX_b-\delta^i_\mu\delta^j_\nu X_iX_j]$ for $AdS_5\times S^5$, $H_{\mu\nu}=-\delta^a_\mu\delta^b_\nu X_aX_b+\delta_\mu^i\delta_\nu^j\left[{1\over4}(X_iX_j-(JX)_i(JX)_j) \right]$ for $AdS_4\times\mathbb{CP}^3$, and $H_{\mu\nu}=-[\delta^a_\mu\delta^b_\nu X_a X_b-\delta^i_\mu\delta^j_\nu p_+^2X_iX_j-\delta^r_\mu\delta^s_\nu p_-^2X_rX_s]$ for $AdS_3\times S^3\times{\cal M}_4$.}. Its form is derived from the diffeomorphism analysis in Appendix \ref{app:Diffeo}, whose explicit expression for our backgrounds and $T^{(0)}$ from \eqref{eqn:LRT0thsol} is given by
\ie
(F_H(T^{(0)}))_{a_1\cdots a_s}=&e^{-iEt}\bigg[\left(-{X_aX^a-\rho_I\over2}+inEt\right)u_{a_1}\cdots u_{a_s}+{(2n+2)z\over2}X_{(a_1}u_{a_2}\cdots u_{a_s)}
\\
&+{in(3n+7)\over4}zk_{(a_1}u_{a_2}\cdots u_{a_s)}\bigg],
\fe
with
\ie
s=2n+2,~~\rho_I=\begin{cases} 
      X_iX^i & \text{for } AdS_5\times S^5 \\
      0  & \text{for } AdS_4\times \mathbb{CP}^3 \\
      p_+^2X_iX^i+p_-^2X_rX^r  &  \text{for } AdS_3\times S^3\times{\cal M}_4
   \end{cases}
\fe
Finally, ${\cal C}_n(T^{(0)})$ stands for mass-level $n$ string fields whose matter and ghost oscillator structures differ from ${\cal V}_n$ that will be required for $\hat\varphi_n^{(2)}$ to solve the linearized EOM.

The linearized EOM with \eqref{eqn:VClinFluc} becomes
\ie\label{eqn:2ndLinEOMRedef}
Q{\cal V}_n[T^{(2)}+F_H(T^{(0)})]+Q{\cal C}_n(T^{(0)})+\hat{S}^{NS}_n+\hat{S}^R_n=0.
\fe
As mentioned before, the field redefinition $F_H(T^{(0)})$ for the leading Regge component of the string field and other components ${\cal C}_n(T^{(0)})$ should be such that \eqref{eqn:2ndLinEOMRedef} leads to the following form of equations for $T^{(2)}$:
\ie\label{eqn:T2equation}
{}&(\Box-4n)T^{(2)}+(\delta\nabla^2)T^{(0)}-M_nT^{(0)}=0,
\\
&\partial\cdot T^{(2)}+(\delta\nabla\cdot)T^{(0)}=0,
\\
&\tr_{\eta}T^{(2)}+(\delta\tr) T^{(0)}=0,
\fe
and this will determine the value of $M_n$. The existence of ${\cal C}_n({T^{(0)}})$ such that this will be the case is shown below by explicitly constructing it. In \eqref{eqn:T2equation}, we introduced $AdS$ deformations of geometric operations at the second order in $\mu$ as
\ie\label{eqn:AdSgeoCorrection}
\nabla_{AdS}^2=\Box+\mu^2\delta\nabla^2+{\cal O}(\mu^4),~~\nabla_{AdS}\cdot=\partial\cdot+\mu^2(\delta\nabla\cdot)+{\cal O}(\mu^4),~~\tr_{AdS}=\tr_{\eta}+\mu^2\delta\tr+{\cal O}(\mu^4),
\fe
whose expressions in the embedding coordinates are presented in \eqref{eqn:AdSOpsecondorderPart}. When acting on ${T}^{(0)}$ of \eqref{eqn:LRT0thsol}, they become
\ie
{}&((\delta\nabla^2){T}^{(0)})_{a_1\cdots a_s}=e^{-iEt}\left[ (-E^2t^2-idEt+s)u_{a_1}\cdots u_{a_s}+2iszk_{(a_1}u_{a_2}\cdots u_{a_s)} \right],
\\
&((\delta\nabla\cdot){T}^{(0)})_{a_2\cdots a_s}=e^{-iEt}(s+d-1-iEt)zu_{a_2}\cdots u_{a_s},
\\
&((\delta\tr){T}^{(0)})_{a_3\cdots a_s}=e^{-iEt}z^2u_{a_3}\cdots u_{a_s}.
\fe

Equations \eqref{eqn:0thEqnSol}, \eqref{eqn:1stOrderEqSol}, \eqref{eqn:T2equation}, and \eqref{eqn:Texpansion} together imply that the $AdS$ wave equation, transversality, and tracelessness \eqref{eqn:AdSlinEOM} are satisfied to the second order in $\mu$. In order for \eqref{eqn:2ndLinEOMRedef} to produce \eqref{eqn:T2equation}, we demand
\ie\label{eqn:linEOMrequired}
{}&Q{\cal V}_n[F_H(T^{(0)})]+Q{\cal C}_n(T^{(0)})+\hat{S}^{NS}_n+\hat{S}^R_n
\\
&={\cal J}_w[(\delta\nabla^2)T^{(0)}-M_nT^{(0)}]+{\cal J}_d[(\delta\nabla\cdot)T^{(0)}]+{\cal J}_t[(\delta\tr)T^{(0)}].
\fe
Note that \eqref{eqn:linEOMrequired} is independent of $T^{(2)}$. The full expressions for $Q{\cal V}_n[F_H(T^{(0)})]$ and ${\cal J}$ terms are presented in Appendix \ref{app:QFH}. \eqref{eqn:linEOMrequired} is now regarded as equations for the unknowns ${\cal C}_n(T^{(0)})$ and $M_n$, and within an appropriate ansatz for ${\cal C}_n(T^{(0)})$ to be discussed below, we will find a solution which fixes $M_n$ to the value given by \eqref{eqn:massCorrection}. Note that one may always add $Q$-closed terms to ${\cal C}_n(T^{(0)})$ without affecting the linearized EOM. To proceed further, we have to evaluate the source terms $\hat{S}^{NS}_n$ and $\hat{S}^R_n$ for specific $AdS$ backgrounds. They depend on $T^{(0)}$ which takes the form \eqref{eqn:LRT0thsol}.

\subsubsection{NSNS source term}
The NSNS source term $\hat{S}^{NS}_n$ is essentially the OPE of $\varphi^{(0)}$ and $\psi^{(2)}$ taken with the local coordinate chart effects dictated by the flat superstring vertex for three-punctured Riemann sphere. The rescaling introduced in \eqref{eqn:rescaling2nd} together with $b$ dependence of $\psi^{(2)}$ in \eqref{eqn:AdS5bgdSol} and \eqref{eqn:AdS4bgdsol}, as shown in \eqref{eqn:bDependence}, cancels against overall $r_0$ and $z_0$ dependence of this flat vertex computation. We introduce the following notations that appear in $\hat{S}^{NS}_n$ (recall that $u^\mu$ is a $\mathbb{R}^{1,9}$ vector defined in \eqref{eqn:LRT0thsol}):
\ie\label{eqn:operatorNotations1}
{}&B_n=ce^{-\phi}(u\cdot\psi)(u\cdot \partial X)^n,~~B_n^0=-ce^{-\phi}\psi^0(u\cdot \partial X)^n,~~U^0_n=-ce^{-\phi} (u\cdot\psi)\partial X^0(u\cdot \partial X)^{n-1},
\\
&F_n=ce^{-\phi}(u\cdot\partial\psi)(u\cdot \partial X)^{n-1},~~Y_n=ce^{-\phi}(u\cdot\psi)(u\cdot\partial^2 X)(u\cdot \partial X)^{n-2},~~G_n=c\eta (u\cdot\partial X)^n,
\\
&P_n=c\partial^2ce^{-\phi}(u\cdot\psi)(u\cdot\partial X)^{n-1},~~A_n=-c\eta(u\cdot\psi)\psi^0(u\cdot\partial X)^{n-1},~~H_n=c\eta(u\cdot\psi)(u\cdot\partial\psi)(u\cdot\partial X)^{n-2}.
\fe
$Y_n$ and $H_n$ are defined only for $n\geq2$. We further define the following linear combinations:
\ie
{}&{\mathscr D}_n=F_n+{n-1\over2}Y_n,~~{\mathscr L}_n={z(diE+8nt)\over8}(B^0_n+nU^0_n)-{inz[2Et+(d+2)i]\over4}{\mathscr D}_n
\\
&{\mathscr K}_n=B^0_n+nU^0_n-{iE\over2}{\mathscr D}_n,~~{\mathscr U}_n={z(2n-d)\over16}G_n+{(d+2)nz\over16}P_n-{iEnz\over8}A_n-{n(n-1)z\over8}H_n.
\fe
Introducing two combinations of holomorphic and anti-holomorphic operators
\ie
A\vee \bar B=A\bar B+B\bar A,~~A\wedge \bar B = A\bar B-B\bar A,
\fe
for any holomorphic operators $A$ and $B$, the source term $\hat S_n^{NS}$ is given by
\ie\label{eqn:NSsourceterm}
{}&\hat S_n^{NS}=N_ne^{-iEt}[c_0^+{\cal P}+{\cal Q}]+2(d-1)n\tau N_ne^{-iEt}c^+_0B_n\bar B_n,
\\
&{\cal P}=\alpha_n(t)B_n\bar B_n+{\mathscr L}_n\vee\bar B_n+{nz^2\over2}{\mathscr K}_n\bar{\mathscr K}_n,~~{\cal Q}={\mathscr U}_n\wedge\bar B_n,
\\
&\alpha_n(t)={1\over4}\left[ 8nt^2-2iE(n-d-1)t-3n^2-5+(d-4)(n+9)+8(d-4)(d-5) \right].
\fe

\subsubsection{RR source term}
The computation of the RR source term $\hat S_n^{R}$ essentially reduces to a certain modified four-point amplitude computation involving two on-shell LRT states $\varphi_n^{(0)}$ and two insertions of RR flux background states $\psi^{(1)}$. This is because $F^{(1)}$ of interest carries zero spin along the $AdS$ directions so that the string bracket of two $F^{(1)}$'s and one LRT state $\varphi_n^{(0)}$ should result in another LRT state which by momentum conservation must also be on-shell. Therefore, we obtain
\ie\label{eqn:RRsourceterm}
\hat S_n^R={\cal Y}N_n e^{-iEt}c_0^+B_n\bar B_n,
\fe
where the number ${\cal Y}$ is determined by
\ie\label{eqn:4ptRR}
{\cal Y}={1\over2}\bigg\langle \tilde\varphi_n^{(0)}\bigg|c_0^-\bigg|\bigg(\left[(\psi^{(1)})^{\otimes2}\otimes\varphi_n^{(0)}\right]-\left[h\left[(\psi^{(1)})^{\otimes2}\right]\otimes\varphi_n^{(0)}\right]-2\left[ \psi^{(1)}\otimes h_n\left[\psi^{(1)}\otimes\varphi_n^{(0)}\right] \right]\bigg)\bigg\rangle_{red},
\fe
with
\ie
\tilde\varphi_n^{(0)}=-{1\over 2^{n+1}n!}e^{iEt}ce^{-\phi}(v\cdot\psi)(v\cdot\partial X)^n\bar ce^{-\bar\phi}(v\cdot\bar\psi)(v\cdot\bar\partial X)^n,~~~v^\mu=(0,1,-i,0,\cdots,0),
\fe
and the reduced correlator $\langle\cdots\rangle_{red}$ is obtained by dividing the ordinary correlator by the momentum-conservation delta function $(2\pi)^{10-d_c}\delta^{10-d_c}(0)V_c$, where $d_c$ and $V_c$ are respectively dimension and volume of the compact directions (we set $V_c=1$ if $d_c=0$). We carry out this computation in Appendix \ref{app:RRsource} which results in
\ie\label{eqn:RRsourceCoeff}
{\cal Y}=-2(d-1)n\tau.
\fe
Therefore, $\hat S^R_n$ term completely cancels the SFT parameter $\tau$-dependent term of $\hat S_n^{NS}$.

\subsubsection{Terms in ${\cal C}_n$}\label{subsubsec:Cn}
The only terms in \eqref{eqn:linEOMrequired} that have not been determined yet are ${\cal C}_n(T^{(0)})$ and $M_n$. We start by writing down an ansatz for ${\cal C}_n(T^{(0)})$. Since other terms in \eqref{eqn:linEOMrequired} carry maximal spin and respect worldsheet parity, so does ${\cal C}_n(T^{(0)})$. We can further restrict ${\cal C}_n(T^{(0)})$ to states that carry no excitations along $3,4,\cdots,9$ directions. By definition, it has zero component along the ${\mathbb R}^{1,9}$ LRT state. However, there could still be states carrying maximal spin because, at this order in $\mu$, one may use up to two powers of the $X$ zero-mode $z=X^1+iX^2$, which carries spin 1 along 1-2 plane, to raise the spin of the string fields. In addition to operators introduced in \eqref{eqn:operatorNotations1}, we further introduce the following notations:
\ie
Z_n=-ce^{-\phi}(u\cdot\psi)\psi^0(u\cdot\partial\psi)(u\cdot\partial X)^{n-2},~~D_n=c e^{-2\phi}\partial\xi(u\cdot\partial X)^n   ,~~L_n=e^{-\phi}(u\cdot\psi)(u\cdot\partial X)^{n-1},
\fe
where $Z_n$ exists only for $n\geq2$. We divide operators into the different sets as
\ie
{\mathfrak U}_n=\{B^0_n,U^0_n,F_n,Y_n,Z_n\},~~{\mathfrak D}_n=\{D_n,L_n\},~~\mathfrak{S}_n=\{G_n,P_n\},~~{\mathfrak U}^0_n=\{B_n^0,U_n^0\},
\fe
For $n\geq2$, five combinations whose spin are less than the maximal spin by 1 are
\ie
{\mathfrak M}^1_n=\{B_n\vee \bar a:a\in{\mathfrak U}_n\setminus{\mathfrak U}_n^0 \}\cup\{c_0^+B_n\wedge\bar d:d\in{\mathfrak D}_n\},
\fe
and 26 combinations whose spin are less than the maximal spin by 2 are
\ie
{\mathfrak M}_n^2=&\{a\bar a:a\in{\mathfrak U}_n\setminus{\mathfrak U}_n^0   \}\cup\{a\vee\bar b:a,b\in{\mathfrak U}_n,a\neq b,(a,b)\notin{\mathfrak U}_n^0\times {\mathfrak U}_n^0\}
\\
&\cup\{g\wedge\bar d:g\in{\mathfrak S}_n,d\in {\mathfrak D}_n\}\cup\{c_0^+a\wedge\bar d:a\in{\mathfrak U}_n,d\in{\mathfrak D}_n\}.
\fe
For $n=1$ case, we remove all operators containing $Y_n$ and $Z_n$ and add $B_1\vee \bar T_1$ (see \eqref{eqn:moreoperators}) to ${\mathfrak M}^1_1$. An ansatz we take for ${\cal C}_n(T^{(0)})$ is
\ie\label{eqn:Cngeneralform}
{\cal C}_n(T^{(0)})=N_ne^{-iEt}\left[ \sum_{{\cal O}\in{\mathfrak M}_n^1}z(a_{\cal O}+b_{\cal O}t){\cal O}+\sum_{{\cal O}\in{\mathfrak M}_n^2}c_{\cal O}z^2{\cal O} \right],
\fe
with in total of 36 independent coefficients $a_{\cal O},b_{\cal O},c_{\cal O}$ for $n\geq2$ and 21 independent coefficients for $n=1$.

\eqref{eqn:linEOMrequired} provides a set of linear equations for these independent coefficients together with $M_n$. One finds a one-parameter family of solutions, where different solutions are gauge-equivalent as they relate to each other by a $Q$-exact term. The value of $M_n$ can be shown to independent of the parameter as expected. Details of this solution are presented in Appendix \ref{app:Cn}.

\subsubsection{Determination of $AdS$ mass correction and comparison against the integrability and $AdS$ Virasoro-Shapiro predictions}
Collecting all contributions, the coefficients multiplying $c_0^+B_n\bar B_n$ in \eqref{eqn:linEOMrequired} are given by
\ie
{}&\underbrace{(-iEt+4n^2-n+9-2d-2(d-4)(d-5))}_{\eqref{eqn:EOMterms1st}:~Q{\cal V}_n[F_H(T^{(0)})]}+\underbrace{\left({iE(n+1)\over2}t-{(n+1)(n+d+3)\over4}\right)}_{\eqref{eqn:CnLRTcomp}:~Q {\cal C}_n(T^{(0)})}
\\
&+\underbrace{\left( 2nt^2+{iE(d+1-n)\over2}t+{-3n^2-5+(d-4)(n+9)+8(d-4)(d-5)\over4}\right)}_{\eqref{eqn:NSsourceterm},~\eqref{eqn:RRsourceterm}:~\hat S_n^{NS}+\hat S_n^R}
\\
&=\underbrace{\left( 2nt^2+{idE\over2}t-(n+1) \right)}_{\eqref{eqn:EOMterms2nd}:~{\cal J}_w[(\delta\nabla^2)T^{(0)}]}+{M_n\over2},
\fe
leading to \eqref{eqn:massCorrection}
\ie
M_n=2(n-1)(3n+1).
\fe
This mass correction holds true for all backgrounds we considered in this work, i.e. $AdS_5\times S^5$, $AdS_4\times\mathbb{CP}^3$, and $AdS_3\times S^3\times{\cal M}_4$ with pure RR flux for both ${\cal M}_4=S^3\times S^1$ and ${\cal M}_4=T^4$. On the dual gauge theory side, this becomes the correction to the conformal dimension in the strong 't Hooft coupling expansion. The dimension $\Delta_{s}$ of the gauge theory operator dual to the LRT state with spin $s=2n+2$ is related to the $AdS$ mass by
\ie
\Delta_s(\Delta_s-d+1)-s=R^2\left[ 4n+{2(n-1)(3n+1)\over R^2}+{\cal O}(R^{-3}) \right],
\fe
from which we obtain
\ie
\Delta_{2n+2}=2\sqrt{n}R+{(d-1)\over2}+{24n^2-8n+(d-1)^2\over16\sqrt{n}R}+{\cal O}(R^{-2}).
\fe

For $AdS_5\times S^5$ case, integrability results for planar ${\cal N}=4$ SYM \cite{Gromov:2011bz,Gromov:2014bva} predict
\ie
\Delta_{\mathfrak{sl}(2)}(S,J)=\sqrt{2S}\lambda^{1/4}+{2J^2+3S^2-2S\over4\sqrt{2S}}\lambda^{-1/4}+{\cal O}(\lambda^{-3/4}),
\fe
as the dimension of the lowest-dimension conformal primary in the $\mathfrak{sl}(2)$ sector at Lorentz spin $S$ and $R$-symmetry charge $J$, with $\lambda$ being the 't Hooft coupling. LRT states in this work correspond to the superconformal descendents of these primaries at $S=2n$ and $J=2$, with $R=\lambda^{1/4}$:
\ie
\Delta_{2n+2}=\Delta_{\mathfrak{sl}(2)}(S=2n,J=2)+2.
\fe

For $AdS_4\times\mathbb{CP}^3$, the quantum spectral curve method for ABJM theory \cite{Gromov:2014eha} predicts
\ie
\left(\Delta_{\mathfrak{sl}(2)}+{1\over2}\right)^2-\left(L+{1\over2}\right)^2=2S\mathfrak{g}-S+{3\over2}S^2+{\cal O}(\mathfrak{g}^{-1}),
\fe
for the dimension $\Delta_{\mathfrak{sl}(2)}$, spin $S$, and $R$-symmetry charge $L$ conformal primary on the even-spin $\mathfrak{sl}(2)$ trajectory, with $\mathfrak{g}$ being the coupling defined in \cite{Gromov:2014eha}. Setting $S=2n$ and $L=1$, with $\mathfrak{g}=R^2$ \cite{Chester:2024esn}, we get the agreement with our LRT states as
\ie
\Delta_{2n+2}=\Delta_{\mathfrak{sl}(2)}+2.
\fe

To the best of our understanding, we are not aware of the integrability or quantum spectral curve results to be compared against our results for $AdS_3\times S^3\times{\cal M}_4$. Instead, recent developments in $AdS$ Virasoro-Shapiro amplitudes provide a direct comparison. For the case ${\cal M}_4=T^4$, \cite{Chester:2024wnb} obtained the poles in the amplitude that translate to the conformal dimensions (we have set $\delta={l\over2}$ in their notation to compare against LRT states)
\ie
\Delta(l)=\sqrt{2l}\lambda^{1/4}-1+{3l^2-2l+2\over4\sqrt{2l}}\lambda^{-1/4}+{\cal O}(\lambda^{-3/4}),
\fe
where $l$ is the Lorentz spin on the boundary. With $l=2n$ and $R=\lambda^{1/4}$, we get the agreement
\ie
\Delta_{2n+2}=\Delta(l=2n)+2.
\fe

\subsection{$AdS$ wavefunction}\label{sec:AdSwave}
We have shown that the tensor $T$ that appears in the the linearized EOM satisfies the expected $AdS$ equations \eqref{eqn:AdSlinEOM} to the second order in $\mu$, with mass correction given by \eqref{eqn:massCorrection}. For completeness, we also show the explicit form of the wavefunction $T$ that solves these equations. Such wavefunctions have previously been constructed in literature, and we mostly follow section 3.1 in \cite{Fitzpatrick:2010zm}.

In the embedding space coordinates we use in this work, it is straightforward to show that the totally symmetric rank-$s$ tensor given by\footnote{This wavefunction can be obtained from (3.3) of \cite{Fitzpatrick:2010zm} by setting $n=0$, $l=s$, $\omega=\Delta$, contracting the lowering Killing vector $\xi_\mu^b$ with our polarization vector $\textbf{u}=(1,i,0,\cdots)$ from $u^a=(0,\textbf{u})$, and lowering the $b$ indices using the metric. Their Killing vector in our coordinates reads $\xi^0_\mu=-iX_\mu,~~\xi^j_\mu=-({\sqrt{R^2+X_aX^a}+it})\delta^j_\mu$ where $\mu,j=1,\cdots,d-1$, leading to $\xi_0\cdot\textbf{u}=z\left(i-{t\over \sqrt{R^2+X_aX^a}}\right)$ and $\xi_j\cdot\textbf{u}=-(\sqrt{R^2+X_aX^a}+it)u_j+{zX_j\over\sqrt{R^2+X_aX^a}}$. Their $(e^{it}\cos\rho)^{\omega+l}$ becomes $q^{-\Delta_s}q^{-s}$ in our notation.}
\ie\label{eqn:AdSWavefunction}
T_{a_1\cdots a_s}(X)=q^{-\Delta_s}\epsilon_{a_1}(X)\cdots\epsilon_{a_s}(X),
\fe
with
\ie
q={\sqrt{R^2+X_aX^a}+it\over R},~~\epsilon_a(X)=u_a-z\partial_a\log q,
\fe
satisfies the $AdS$ equations
\ie
\left(\nabla_{AdS}^2-{\Delta_s(\Delta_s-d+1)-s\over R^2}\right) T=0,~~\nabla_{AdS}\cdot T=0,~~\tr_{AdS} T=0.
\fe
Expanding in small $\mu=R^{-1}$, we obtain \eqref{eqn:LRT0thsol} for $T^{(0)}$ and \eqref{eqn:LRT1stTsol} for $T^{(1)}$ as expected. This is the reason why we chose the first order solution $\varphi^{(1)}_n$ to be \eqref{eqn:LRT1stsol} as opposed to zero for example. Even though this choice did not affect the linearized EOM at order $\mu^2$, we expect that it will affect higher order solutions. Finally, a solution $T^{(2)}$ to \eqref{eqn:T2equation} can be obtained simply by expanding \eqref{eqn:AdSWavefunction} to the second order in $\mu$ and taking order $\mu^2$ term.

\section{Conclusion}
In this work, we demonstrated that SFT provides a computational framework for deriving physical observables in holographic string backgrounds in $AdS$. The first curvature corrections to the spectrum of LRT states for $AdS_5\times S^5$ and $AdS_4\times\mathbb{CP}^3$ computed in this work reproduce the integrability and quantum spectral curve predictions of the dual CFTs. Given that integrability and quantum spectral curve methods are based on a series of conjectures, SFT results in this work can also be regarded as a nontrivial check of such methods. We conclude by discussing suggestive future directions.

Firstly, it is straightforward to extend the background solutions and linearized EOM to higher orders in the small curvature expansion. Such an extension requires higher-point sphere correlators of the worldsheet CFT, which could become computationally intense. However, recent developments in computationally efficient string vertices \cite{Mazel:2024alu}, code packages such as \StringCode\ \cite{StringCode} that automatize such computations, and large language models that write such codes very efficiently, make the acquirement of higher-order results much more feasible than before. Such results will provide highly nontrivial precision tests of the holographic duality.

SFT also allows for the computation of other observables in $AdS$ backgrounds. SFT action expanded around the $AdS$ solution provides Feynman rules for computing genuinely stringy observables in $AdS$ beyond just the spectrum. Dual gauge theory predictions on many of such observables are currently not available, and SFT results therefore provide intriguing predictions on the dual gauge theory. For example, SFT provides first-principles rules for computing the $AdS$ Virasoro-Shapiro amplitudes that have been intensively studied in recent years \cite{Alday:2023jdk,Alday:2023mvu}. One key difference of SFT from other approaches to such observables is that there are no assumptions or guesses involved in SFT computations, which makes its results free of ambiguities.

Another further extension is to study higher genus corrections to stringy observables in $AdS$ which correspond to ${1\over N}$ corrections. SFT framework is fully consistent at the quantum level order by order in the genus expansion and thus offers methods to compute observables at higher orders in the string coupling. Practical difficulties concerning the higher genus correlators of the worldsheet CFT and nontrivial roles of PCOs and vertical integrations are present, but there also have been encouraging recent developments in the computation of higher genus observables in perturbative string theory \cite{Eberhardt:2023xck,Baccianti:2025gll,Baccianti:2025whd,Christian:2026maj}.

Finally, SFT provides a perturbative method to study string backgrounds other than $AdS$, such as flux compactification as studied in \cite{Cho:2023mhw,Kim:2024dnw}. These works established the relevant SFT setup but did not compute observables beyond those predicted by supergravity. Observables such as $\alpha'$ corrections to the K\"ahler potential provide significant information about such backgrounds that was previously unavailable and deserves serious investigations.

\section*{Acknowledgments}
We would like to thank Shai Chester, Shota Komatsu, Rodrigo S. Pitombo, Bogdan Stefanski, and Barton Zwiebach for enlightening discussions. MC, JS, and XY thank the RIKEN Interdisciplinary Theoretical and Mathematical Sciences Program (iTHEMS) and the Yukawa Institute for Theoretical Physics (YITP) at Kyoto University for their hospitality during the iTHEMS-YITP workshop 'Resurgence, Matrix model, String field theory and Instantons' / 'Really Marvelous Sparkling Ideas' (YITP-W-26-11), during the course of this work. JG thanks the Centro de Ciencias Pedro Pascual during the workshop 'String field theory and flux compactification', the Institut d'Etudes Scientifiques de Cargèse during the 'Advanced Summer School in Quantum Field Theory and Quantum Gravity', and ICTS during the program 'Worldsheet String Theory and String Field Theory', where parts of this work were completed, for their hospitality. The work of MC is supported by the Simons Collaboration on Global Categorical Symmetries. The work of JG is supported by FAPESP grants 2022/04105-0 and 2024/00082-1.

We acknowledge the use of ChatGPT 5.6, 6.0, and 6.1 (OpenAI) and Opus 5.5 (Anthropic) in performing SFT computations, in particular the worldsheet CFT computations, which were independently verified using \StringCode~\cite{StringCode} at mass-levels $n=1,2$. In \cite{StringCodeAdSMassShifts}, we have posted a \texttt{Mathematica} notebook \texttt{AdSLRTmasses.nb} that presents \StringCode~computations for this work.

\appendix
\section{Spinor and worldsheet conventions}\label{app:spinor}
We briefly review the spinor and worldsheet CFT conventions used in the work, which follow those used in \cite{Cho:2025coy}. Throughout the text, we denote the chiral spin field by $S_{\alpha}$ and the anti-chiral spin field by $S^{\alpha}$. These can be assembled into a Dirac spinor
\begin{equation}
    S_A = \begin{pmatrix}
        S_{\alpha}\\
        S^{\alpha}
    \end{pmatrix}.
\end{equation}
Since in type IIA(B) the left-moving R-sector has opposite (same) chirality with respect to its right-moving counterpart, following these conventions, the RR-sector string fields considered above are taken to be (anti-chiral,chiral) in type IIA and (anti-chiral,anti-chiral) in type IIB. In both cases, they will be taken in the $-\frac{1}{2}$ picture. In this basis, the gamma-matrices take the off-diagonal form
\begin{equation}
    (\Gamma^{\mu})_A\,^B=
    \begin{pmatrix}
        0 & (L^{\mu})_{\alpha\beta}\\
        (U^{\mu})^{\alpha\beta} & 0
    \end{pmatrix}.
\end{equation}
The $16\times 16$ off-diagonal matrices $(L^{\mu})_{\alpha\beta},~ (U^{\mu})^{\alpha\beta}$ have real entries, are symmetric in $\alpha,\beta$ and satisfy the Clifford algebra
\ie
L^\mu U^\nu+L^\nu U^\mu=2\eta^{\mu\nu},~~~U^\mu L^\nu+U^\nu L^\mu = 2\eta^{\mu\nu}.
\fe
The charge conjugation matrix $C=\Gamma^0$ and its inverse are given by
\begin{equation}
    C_A\,^B = \begin{pmatrix}
        0 & \delta_{\alpha}\,^{\beta}\\
        -\delta^{\alpha}\,_{\beta} & 0
    \end{pmatrix},\quad
    (C^{-1})^{A}_{~B} = \begin{pmatrix}
        0 & -\delta_{\alpha}\,^{\beta}\\
        \delta^{\alpha}\,_{\beta} & 0
    \end{pmatrix}.
\end{equation}
Likewise, the chirality operator $\Gamma \equiv \Gamma^0\Gamma^1\cdots\Gamma^9$ is expressed as
\begin{equation}
    \Gamma_A\,^{B} = \begin{pmatrix}
        -\delta_{\alpha}\,^{\beta} & 0\\
        0 & \delta^{\alpha}\,_{\beta}
    \end{pmatrix},
\end{equation}
while the chiral and anti-chiral projectors $P_{\mp}=\frac{1}{2}\left(1\mp\Gamma\right)$ read
\begin{equation}
    (P_+)_A\,^B = \begin{pmatrix}
        0 & 0\\
        0 & \delta^{\alpha}\,_{\beta}
    \end{pmatrix},\quad
    (P_-)_A\,^B = \begin{pmatrix}
        \delta_{\alpha}\,^{\beta} & 0\\
        0 & 0
    \end{pmatrix},
\end{equation}
such that $P_+S_A = S^{\alpha}$ and $P_-S_A = S_{\alpha}$. Finally, note that
\begin{equation}
    (\Gamma^{\mu}C)_A\,^B = 
    \begin{pmatrix}
        -(L^{\mu})_{\alpha\beta} & 0\\
        0 & (U^{\mu})^{\alpha\beta}
    \end{pmatrix},\quad
    (C^{-1}\Gamma^{\mu})_A\,^B = 
    \begin{pmatrix}
        -(U^{\mu})^{\alpha\beta} & 0\\
        0 & (L^{\mu})_{\alpha\beta}
    \end{pmatrix}.
\end{equation}

In these conventions, the OPEs we use in our computations take the following form
\begin{equation}
\begin{split}
    \text{Type IIA/B:}\quad e^{-\frac{1}{2}\phi}S^{\alpha}(z)e^{-\frac{1}{2}\phi}S^{\beta}(w)\sim-i(U_{\mu})^{\alpha\beta}\frac{e^{-\phi}\psi^{\mu}(w)}{z-w}+\cdots,\\
    \text{Type IIA:}\quad e^{-\frac{1}{2}\bar\phi}\bar S_{\alpha}(\bar z)e^{-\frac{1}{2}\bar\phi}\bar S_{\beta}(\bar w)\sim+i(L_{\mu})_{\alpha\beta}\frac{e^{-\bar\phi}\bar\psi^{\mu}(\bar w)}{\bar z-\bar w}+\cdots,\\
    \text{Type IIB:}\quad e^{-\frac{1}{2}\bar\phi}\bar S^{\alpha}(\bar z)e^{-\frac{1}{2}\bar\phi}\bar S^{\beta}(\bar w)\sim-i(U_{\mu})^{\alpha\beta}\frac{e^{-\bar\phi}\bar\psi^{\mu}(\bar w)}{\bar z-\bar w}+\cdots,\\
    \text{Type IIA/B:}\quad e^{-\phi}\psi^\mu(z)e^{-\frac{1}{2}\phi}S^\alpha(w) \sim +\frac{i}{2}\frac{(U^\mu)^{\alpha\beta}}{(z-w)}e^{-\frac{3}{2}\phi}S_\beta(w)+\cdots,\\
    \text{Type IIA:}\quad e^{-\bar\phi}\bar\psi^\mu(\bar z)e^{-\frac{1}{2}\bar\phi}\bar S_\alpha(\bar w) \sim -\frac{i}{2}\frac{(L^\mu)_{\alpha\beta}}{(\bar z-\bar w)}e^{-\frac{3}{2}\bar\phi}\bar S^\beta(\bar w)+\cdots,\\
    \text{Type IIB:}\quad e^{-\bar\phi}\bar\psi^\mu(\bar z)e^{-\frac{1}{2}\bar\phi}\bar S^\alpha(\bar w) \sim +\frac{i}{2}\frac{(U^\mu)^{\alpha\beta}}{(\bar z-\bar w)}e^{-\frac{3}{2}\bar\phi}\bar S_\beta(\bar w)+\cdots,\\
    \text{Type IIA/B:}\quad e^{+\phi}\psi^\mu(z)e^{-\frac{1}{2}\phi}S^\alpha(w) \sim +\frac{i}{2}(U^\mu)^{\alpha\beta}e^{+\frac{1}{2}\phi}S_\beta(w)+\cdots,\\
    \text{Type IIA:}\quad e^{+\bar\phi}\bar\psi^\mu(\bar z)e^{-\frac{1}{2}\bar\phi}\bar S_\alpha(\bar w) \sim -\frac{i}{2}(L^\mu)_{\alpha\beta}e^{+\frac{1}{2}\bar\phi}\bar S^\beta(\bar w)+\cdots,\\
    \text{Type IIB:}\quad e^{+\bar\phi}\bar\psi^\mu(\bar z)e^{-\frac{1}{2}\bar\phi}\bar S^\alpha(\bar w) \sim +\frac{i}{2}(U^\mu)^{\alpha\beta}e^{+\frac{1}{2}\bar\phi}\bar S_\beta(\bar w)+\cdots,\\
    \text{Type IIA/B:}\quad e^{+\phi}\psi^\mu(z)e^{-\frac{3}{2}\phi}S_\alpha(w) \sim -\frac{i}{2}(z-w)(L^\mu)_{\alpha\beta}e^{-\frac{1}{2}\phi}S^\beta(w)+\cdots\\
    \text{Type IIA:}\quad e^{+\bar\phi}\bar\psi^\mu(\bar z)e^{-\frac{3}{2}\bar\phi}\bar S^\alpha(\bar w) \sim +\frac{i}{2}(\bar z-\bar w)(U^\mu)^{\alpha\beta}e^{-\frac{1}{2}\bar\phi}\bar S_\beta(\bar w)+\cdots\\
    \text{Type IIB:}\quad e^{+\bar\phi}\bar\psi^\mu(\bar z)e^{-\frac{3}{2}\bar\phi}\bar S_\alpha(\bar w) \sim -\frac{i}{2}(\bar z-\bar w)(L^\mu)_{\alpha\beta}e^{-\frac{1}{2}\bar\phi}\bar S^\beta(\bar w)+\cdots\\
    \text{Type IIA/B:}\quad e^{+\phi}\psi^{\mu}(z)e^{-\frac{5}{2}\phi}S_{\alpha}(w)\sim+\frac{i}{2}(z-w)^2(L^\mu)_{\alpha\beta}e^{-\frac{3}{2}\phi}S^\beta(w)+\cdots\\
    \text{Type IIA:}\quad e^{+\bar\phi}\bar\psi^\mu(\bar z)e^{-\frac{5}{2}\bar\phi}\bar S^\alpha(\bar w) \sim -\frac{i}{2}(\bar z-\bar w)^2(U^\mu)^{\alpha\beta}e^{-\frac{3}{2}\bar\phi}\bar S_\beta(\bar w)+\cdots\\
    \text{Type IIB:}\quad e^{+\bar\phi}\bar\psi^\mu(\bar z)e^{-\frac{5}{2}\bar\phi}\bar S_\alpha(\bar w) \sim +\frac{i}{2}(\bar z-\bar w)^2(L^\mu)_{\alpha\beta}e^{-\frac{3}{2}\bar\phi}\bar S^\beta(\bar w)+\cdots.
\end{split}
\end{equation}
The correlators, on the other hand, are given by
\begin{equation}
\begin{split}
 \text{Type IIA/B:}\quad \big\langle e^{-\phi}\psi^\mu(z_1)
        e^{-\phi/2}S^\alpha(z_2)e^{-\phi/2}S^\beta(z_3)\big\rangle
 =-\frac{i}{2}\frac{ (U^\mu)^{\alpha\beta}}{z_{12}z_{13}z_{23}},\\
 \text{Type IIA:}\quad \big\langle e^{-\bar\phi}\bar\psi^\mu(\bar z_1)
        e^{-\frac{1}{2}\bar\phi}\bar S_\alpha(\bar z_2)e^{-\frac{1}{2}\bar\phi}\bar S_\beta(\bar z_3)\big\rangle
 =+\frac{i}{2}\frac{ (L^\mu)_{\alpha\beta}}{\bar z_{12}\bar z_{13}\bar z_{23}},\\
 \text{Type IIB:}\quad \big\langle e^{-\bar\phi}\bar\psi^\mu(\bar z_1)
        e^{-\frac{1}{2}\bar\phi}\bar S^\alpha(\bar z_2)e^{-\frac{1}{2}\bar\phi}\bar S^\beta(\bar z_3)\big\rangle
 =-\frac{i}{2}\frac{ (U^\mu)^{\alpha\beta}}{\bar z_{12}\bar z_{13}\bar z_{23}},
\end{split}
\end{equation}
where $z_{ij}\equiv z_i-z_j$ and $\bar z_{ij}\equiv \bar z_i-\bar z_j$ and from which we can obtain 4-point correlators by Ward identities as in Appendix \ref{app:RRsource}. In order to go from the conventions above to the \StringCode\ conventions which we have used to verify computations, we Wick rotate the vector indices to Euclidean signature by $\Gamma^0\to -i\Gamma^0$, such that the identification $\Gamma^0=C$ no longer holds. Indeed, there is a non-trivial map between the spin-fields above and those of \StringCode. To relate the two, make $C_{\texttt{sc}}= -iC_{\mathrm{old}}$ and use
\begin{equation}
\begin{aligned}
(e^{-\phi/2}S^a)(z)
&=\frac{
\texttt{S}(s_{w(a)},-\tfrac{1}{2};z)
-i\,\texttt{S}(s_a,-\tfrac{1}{2};z)}
{\sqrt{2}},
\\[4pt]
(e^{-\phi/2}S_a)(z)
&=\frac{
-d_a\,\texttt{S}(-s_a,-\tfrac{1}{2};z)
+i\,d_{w(a)}\,\texttt{S}(-s_{w(a)},-\tfrac{1}{2};z)}
{\sqrt{2}},
\\[4pt]
(e^{-3\phi/2}S^a)(z)
&=\frac{
-d_{w(a)}\,\texttt{S}(s_{w(a)},-\tfrac{3}{2};z)
+i\,d_a\,\texttt{S}(s_a,-\tfrac{3}{2};z)}
{\sqrt{2}},
\\[4pt]
(e^{-3\phi/2}S_a)(z)
&=\frac{
-\texttt{S}(-s_a,-\tfrac{3}{2};z)
+i\,\texttt{S}(-s_{w(a)},-\tfrac{3}{2};z)}
{\sqrt{2}},
\\[4pt]
(e^{-5\phi/2}S^a)(z)
&=\frac{
-i\,\texttt{S}(s_{w(a)},-\tfrac{5}{2};z)
-\texttt{S}(s_a,-\tfrac{5}{2};z)}
{\sqrt{2}},
\\[4pt]
(e^{-5\phi/2}S_a)(z)
&=\frac{
-i\,d_a\,\texttt{S}(-s_a,-\tfrac{5}{2};z)
-d_{w(a)}\,\texttt{S}(-s_{w(a)},-\tfrac{5}{2};z)}
{\sqrt{2}},
\\[4pt]
(e^{+\phi/2}S^a)(z)
&=\frac{
-d_{w(a)}\,\texttt{S}(s_{w(a)},+\tfrac{1}{2};z)
+i\,d_a\,\texttt{S}(s_a,+\tfrac{1}{2};z)}
{\sqrt{2}}.
\end{aligned}
\end{equation}
where $\texttt{S}(s,p;z)$ stands for the \StringCode\ spinor-field as in
\begin{gather*}
\texttt{S}(s,p;z)
 \equiv \texttt{S}\big[\{s,\chi(s)\},p,\{\},0,z\big],
\qquad
\chi(s_a)=\texttt{"chiral"},\quad
\chi(-s_a)=\texttt{"antichiral"},\\
\begin{array}{c|rrrrrrrrrrrrrrrr}
a    & 1 & 2 & 3 & 4 & 5 & 6 & 7 & 8 & 9 & 10 & 11 & 12 & 13 & 14 & 15 & 16\\
\hline
w(a) &16 &15 &14 &11 &13 &10 & 9 &12 & 7 & 6  & 4  & 8  & 5  & 3  & 2  & 1
\end{array}\\
(s_1,\ldots,s_{16})
 \equiv \texttt{spinProjectionSpinBasisState}[\texttt{"chiral"}],\\
d_a=
\begin{cases}
+1,&a\in\{1,12,\ldots,16\},\\
-1,&a\in\{2,\ldots,11\}.
\end{cases}
\end{gather*}

\section{Flat superstring vertex}\label{app:flatV}
In this appendix, we review the flat superstring vertex introduced in \cite{Mazel:2024alu,Cho:2025coy}. It defines a string frame where the computation of string brackets is simplified while preserving the $L_{\infty}$ structure that guarantees a consistent gauge-invariant string field equation. Naturally, this frame is related to the more standard symmetric frames by a field redefinition.

The asymmetry lies in the different treatment of the ``outgoing'' and ``incoming'' coordinate maps. While the former involves inversion, the latter are defined in terms of scaling and translation alone. More precisely, for the $(n+1)$-vertex we have
\begin{equation} \label{eqn:flatCoordMaps}
    f_1(w_1) = r_0 w_1^{-1},\quad f_{i+1}(w_{i+1})=q_iw_{i+1}+z_{i},\quad i=1,\cdots,n.
\end{equation}
These maps determine the ``flat'' superstring bracket via
\begin{equation}
    [\Psi^{\otimes n}] = \mathcal{G}\frac{b_0^-\mathbb{P}^-r_0^{-L_0^+}}{(-2\pi i)^{n-2}}\int_{\Upsilon_{n;1}}e^{\mathcal{B}}\prod_{a=1}^{d_o}\left(\mathcal{X}(x_a)-d\xi(x_a)\right)\prod_{\bar{a}=1}^{\bar{d}_o}\left(\bar{\mathcal{X}}(\bar{x}_{\bar{a}})-d\bar{\xi}(\bar{x}_{\bar{a}})\right)\prod_{i=1}^n\left(|q_i|^{L_0^+}\Psi_i\right)(z_i),
\end{equation}
where $\mathbb{P}^-$ denotes the projection onto the subspace of level-matched states, and $\mathcal{B}$ is given by
\begin{equation}
    \mathcal{B} = -\sum_{i=1}^n\left(dz_ib_{-1}^{(z_i)}+\frac{dq^i}{q_i}b_0^{(z_i)}+d\bar{z}_i\bar{b}_{-1}^{(\bar{z}_i)}+\frac{d\bar{q}^i}{\bar{q}_i}\bar{b}_0^{(\bar{z}_i)}\right).
\end{equation}
The integration contour is performed over $\Upsilon_{n;1}$, a $(2n-4)$-dimensional chain in $\hat{Q}_{n+1}\to\mathcal{M}_{0,n+1}$ specified by the coordinate maps (\ref{eqn:flatCoordMaps}) with a symmetric choice of $(q_i,z_i)$ as functions of the moduli domain $\pi(\Upsilon_{n;1})\subset\mathcal{M}_{0,n+1}$, in addition to the holomorphic and anti-holomorphic PCO positions denoted by $x_1,\cdots,x_{d_o}$ and $\bar{x}_1,\cdots,\bar{x}_{\bar{d}_o}$ respectively. Additionally, these chains also satisfy the matching conditions
\begin{equation}
    -\partial\Upsilon_{0,n;1} = \sum_{\alpha\subset\{1,\cdots,n\}}\varrho_{\alpha}\left(\widetilde{\Upsilon}_{0,|\alpha|;1}\times\widetilde{\Upsilon}_{0,n-|\alpha|+1;1}\times\{q:|q|=1\}\right)
\end{equation}
where $\varrho_{\alpha}$ is the plumbing map such that the sets $(q_i',z_i',x_{a'}',\bar x'_{\bar a'})\in \Gamma_{0,|\alpha|;1}$ and $(q_i'',z_i'',x_{a''}'',,\bar x''_{\bar a''})\in \Gamma_{0,n-|\alpha|+1;1}$ are taken to
\begin{equation}
\begin{split}
   &z_{\alpha(i)}=r_0^{-1}qq_1''z_i'+z_1'',\quad |q_{\alpha(i)}|=r_0^{-1}|q_1''q_i'|,\quad i=1,\ldots,|\alpha|\\
    &z_{\bar \alpha(j)}=z_{j+1}'',\quad |q_{\bar \alpha(j)}|=|q_{j+1}''|,\quad j=1,\ldots,n-|\alpha|,\\
    &x_{a'}=r_0^{-1}qq_1''x'_{a'}+z_1'',\quad \bar x_{\bar a'}=r_0^{-1}\bar q\bar q_1''\bar x'_{\bar a'}+\bar z_1'',\quad x_{a''}=x''_{a''},\quad \bar x_{\bar a''}=\bar x''_{\bar a''}
\end{split}.
\end{equation}
In practice, the flat 2-string bracket takes the simple form
\begin{equation}
    [\Psi^{\otimes 2}] = \mathcal{G}^{\#}[\Psi^{\otimes 2}]_b
\end{equation}
where $\mathcal{G}^{\#}$ denotes the necessary PCO action to achieve $-1$ picture in the NS sector and $-\frac{1}{2}$ picture in the R sector, and
\begin{equation}
    [\Psi^{\otimes 2}]_b \equiv b_0^-\mathbb{P}^-r_0^{-L_0^+}\left(\Psi(-z_0)\Psi(z_0)\right)
\end{equation}
where the field product is understood as an OPE.

The moduli space decomposition for the 4-punctured sphere will be important in the computation of the flux contributions at order $\frac{1}{R^2}$. Let us start by considering two cubic vertices using the coordinate maps of Equation (\ref{eqn:flatCoordMaps})
\begin{gather} \label{eqn:flatCoordMapr0z0}
    z_1 = f_1(w_1) = r_0w_1^{-1},\quad z_1=f_2(w_2)=w_2-z_0,\quad z_1 = f_3(w_3) = w_3+z_0,\\
    z_2 = f_1(w'_1) = r_0w'_1\,^{-1},\quad z_2=f_2(w'_2)=w'_2-z_0,\quad z_2 = f_3(w'_3) = w'_3+z_0. \label{eqn:flatCoordMapr0z0p}
\end{gather}
We will take the two plumbing parameters to be defined by
\begin{equation}
    w'_1 w_2 = q\quad\mathrm{and}\quad w'_1w_3 = q'
\end{equation}
with $|q|\leq1,|q'|\leq1$. In $z_1$'s coordinate frame, the points $z_2=\pm z_0$ are taken to $z_1=\left(\pm\frac{q}{r_0}-1\right)z_0$. By an $SL(2,\mathbb{C})$ transformation, the points $z_1=\infty,z_0, \left(\pm\frac{q}{r_0}-1\right)z_0$ can be mapped to $z_1 = 0,1,t,\infty$, with modulus $t=\frac{2q}{q-2r_0}$. The corresponding propagator region $|q|\leq1$ is therefore given by
\begin{equation}
    \left|\frac{1}{t}-\frac{1}{2}\right|\geq r_0.
\end{equation}
Analogous analysis of the other propagator channels determines the fundamental vertex region
\begin{equation}
    \mathcal{D}_{0,4}=\left\{t\in\mathbb{C}:\left|t-\frac{1}{2}\right|,\left|\frac{1}{t}-\frac{1}{2}\right|,\left|\frac{1}{1-t}-\frac{1}{2}\right|< r_0\right\}.
\end{equation}
We can then write down the definition for the flat 3-bracket as
\begin{equation} \label{eqn:3bracket}
    [\Psi^{\otimes 3}]=\mathcal{G}^{\#}\frac{b_0^-\mathbb{P}^-r_0^{-L_0^+}}{(-2\pi i)}\int_{\mathcal{D}_{0,4}}dt\wedge d\bar{t}~\mathcal{B}_t\bar{\mathcal{B}}_{\bar{t}}\prod_{i=1}^3\left(|q_i|^{L_0^+}\Psi\right)(z_i)-3\mathrm{Vert}\left[[\Psi^{\otimes2}]_b\otimes\Psi\right]
\end{equation}
and the vertical integration contribution $\mathrm{Vert}[\Phi\otimes\Psi]$ is zero for a string field $\Phi$ carrying picture $-1$ or $-\frac{1}{2}$. If $\Phi$ has holomorphic picture number $-2$ or $-\frac{3}{2}$ and anti-holomorphic picture number $-1$ or $-\frac{1}{2}$, then
\begin{equation}
    \mathrm{Vert}[\Phi\otimes\Psi]=\mathcal{G}^{\#}\left(\left[\xi_0\bar{\mathcal{X}}_0\Phi\otimes\Psi\right]_b-\xi_0\left[\bar{\mathcal{X}}_0\Phi\otimes\Psi\right]_b\right),
\end{equation}
whereas if $\Phi=[\Psi^{\otimes 2}]_b$ has holomorphic picture number $-2$ or $-\frac{3}{2}$ and anti-holomorphic picture number $-2$ or $-\frac{3}{2}$, then
\begin{equation}
    \mathrm{Vert}[\Phi\otimes\Psi]=\mathcal{G}^{\#}\left(\left[\xi_0\bar{\mathcal{X}}_0\Phi\otimes\Psi\right]_b-\xi_0\left[\bar{\mathcal{X}}_0\Phi\otimes\Psi\right]_b+\mathcal{X}_0\left[\bar{\xi}_0\Phi\otimes\Psi\right]_b-\mathcal{X}_0\bar{\xi}_0\left[\Phi\otimes\Psi\right]_b\right).
\end{equation}
Note that Equation (\ref{eqn:3bracket}) uses a different sign convention than \cite{Cho:2025coy}, but agrees with the one originally proposed in \cite{Mazel:2024alu}. In particular, it implies a minus sign in integrated vertex operators, such as the one used in (\ref{eqn:4ptCorr}).

Finally, the $r_0$-dependence of the string field in the flat bracket frame must be consistent with
\begin{equation}
    \partial_{r_0}\Psi(r_0) = \sum_{n=2}^{\infty}\frac{1}{n!}[[\Psi^{\otimes n}]]_{r_0}
\end{equation}
where $[[\Psi^{\otimes n}]]_{r_0}$ denotes the Hata-Zwiebach double-bracket \cite{Hata:1993gf}. It is defined as
\begin{equation}\label{eqn:hataZwiebach}
    [[\Psi^{\otimes n}]]_{r_0} = \mathcal{G}\frac{b_0^-\mathbb{P}^-r_0^{-L_0^+}}{(-2\pi i)^{n-2}}\int_{\Upsilon_{n;1}}\mathcal{B}_{r_0}e^{\mathcal{B}}\prod_{a=1}^{d_o}\left(\mathcal{X}(x_a)-d\xi(x_a)\right)\prod_{\bar{a}=1}^{\bar{d}_o}\left(\bar{\mathcal{X}}(\bar{x}_{\bar{a}})-d\bar{\xi}(\bar{x}_{\bar{a}})\right)\prod_{i=1}^n\left(|q_i|^{L_0^+}\Psi_i\right)(z_i),
\end{equation}
where the extra $b$-ghost insertion $\mathcal{B}_{r_0}$ measures the coordinate maps' dependence on $r_0$. It will be convenient for our purposes that the 2-bracket reduces to
\begin{equation}
    [[\Psi^{\otimes 2}]]_{r_0} = \frac{1}{r_0}b_0^+[\Psi^{\otimes 2}]_b.
\end{equation}

\section{$AdS$ supergravity backgrounds}\label{app:AdSsugra}
In this appendix, we review the supergravity solutions describing the $AdS$ backgrounds of interest.

\subsection{$AdS_5\times S^5$}
The geometry of the $AdS_5\times S^5$ can be embedded in $\mathbb{R}^{2,10}$ via
\begin{gather}
    -(x^{-1})^2-(x^0)^2+(x^1)^2+\cdots+(x^4)^2=-R^2,\\
    (x^{5})^2+(x^6)^2+(x^7)^2+\cdots+(x^{10})^2=+R^2.
\end{gather}
The \textit{embedding coordinates} $\{x^a,x^i\}$ with $a=0,\cdots,4$ and $i=5,\cdots,9$ then satisfy
\begin{equation}
    x^{-1} = \pm\sqrt{R^2+x_ax^a},\quad x^{10}=\pm\sqrt{R^2-x_ix^i}.
\end{equation}
In these coordinates, the metric takes the form
\begin{gather}
    G_{ab} = \eta_{ab}-\frac{x_ax_b}{R^2+x_ax^a} = \eta_{ab}-\frac{x_ax_b}{R^2}+O(R^{-4}),\\
    G_{ij} = \delta_{ij}+\frac{x_ix_j}{R^2-x_ix^i} = \delta_{ij}+\frac{x_ix_j}{R^2}+O(R^{-4}).
\end{gather}
The background is supported by the 5-form RR flux
\begin{equation}
    F = \frac{1}{R}(\mathrm{vol}_{AdS_5}+\mathrm{vol}_{S^5}),
\end{equation}
which, in a $1/R$ expansion, reads
\begin{equation}
    F = \frac{1}{R}\left(1-\frac{1}{2R^2}x_ax^a\right)dx^0\wedge\cdots\wedge dx^4+\frac{1}{R}\left(1+\frac{1}{2R^2}x_ix^i\right)dx^5\wedge\cdots\wedge dx^9+O(R^{-5}).
\end{equation}

\subsection{$AdS_4\times\mathbb{CP}^3$}
Coordinates for $AdS_4$ are given by $a=0,\cdots,3$ and those for $\mathbb{CP}^3$ by $i=4,\cdots,9$. For the $AdS_4$ part, we use the embedding coordinates similar to the $AdS_5$ case:
\ie
G_{ab}=\eta_{ab}-{x_ax_b\over R^2+x_ax^a}=\eta_{ab}-{x_ax_b\over R^2}+{\cal O}(R^{-4}).
\fe
For the $\mathbb{CP}^3$, we work in the following coordinates where K\"ahler form $J$ is taken to be $J_{45}=J_{67}=J_{89}=1$ with other independent components being zero:
\ie
G_{ij}=\delta_{ij}+{x_ix_j\over 4R^2-x_ix^i}-{(Jx)_i(Jx)_j\over4R^2}=\delta_{ij}+{x_ix_j-(Jx)_i(Jx)_j\over 4R^2}+{\cal O}(R^{-4}).
\fe
RR field strengths for this background are
\ie
{}&F_2={1\over R}J,~~~F_4=3\omega_{AdS_4}dx^0\wedge dx^1\wedge dx^2\wedge dx^3,
\\
&\omega_{AdS_4}={1\over R}\left(1+{1\over R^2}x_ax^a \right)^{-1/2}={1\over R}\left(1-{1\over2R^2}x_ax^a+{\cal O}(R^{-4})\right).
\fe

\subsection{$AdS_3\times S^3\times{\cal M}_4$}
We first present the case of ${\cal M}_4=S^3\times S^1$. The \textit{embedding coordinates} $\{x^a,x^i,x^r\}$ are divided into $a=0,1,2$ for $AdS_3$, $i=3,4,5$ for $S^3$, and $r=6,7,8$ for another $S^3$, each with radii $R,R_+,$ and $R_-$ ($S^1$ along the 9-th direction is not displayed below). They satisfy
\ie
{1\over R^2}={1\over R_+^2}+{1\over R_-^2},
\fe
and we introduce $p_\pm={R\over R_\pm}$. In these coordinates, the metric takes the form
\begin{gather}
    G_{ab} = \eta_{ab}-\frac{x_ax_b}{R^2+x_ax^a} = \eta_{ab}-\frac{x_ax_b}{R^2}+O(R^{-4}),\\
    G_{ij} = \delta_{ij}+\frac{x_ix_j}{R_+^2-x_ix^i} = \delta_{ij}+p_+^2\frac{x_ix_j}{R^2}+O(R^{-4}),\\
    G_{rs} = \delta_{rs}+\frac{x_rx_s}{R_-^2-x_rx^r} = \delta_{rs}+p_-^2\frac{x_rx_s}{R^2}+O(R^{-4})
\end{gather}
The background carries the 3-form RR flux
\begin{equation}
    F = \frac{2}{R}(\mathrm{vol}_{AdS_3}+p_+\mathrm{vol}_{S^3_+}+p_-\mathrm{vol}_{S^3_-}),
\end{equation}
where $S^3_{\pm}$ are $S^3$ with radii $R_\pm$. In the $1/R$ expansion, these become
\ie
    F = \frac{2}{R}\bigg[&\left(1-\frac{1}{2R^2}x_ax^a\right)dx^0\wedge dx^1\wedge dx^2+p_+\left(1+\frac{p_+^2}{2R^2}x_ix^i\right)dx^3\wedge dx^4\wedge dx^5
    \\
    &+p_-\left(1+\frac{p_-^2}{2R^2}x_rx^r\right)dx^6\wedge dx^7\wedge dx^8\bigg]+O(R^{-5}).
\fe

\subsection{Order $R^{-2}$ part of $AdS$ geometric operations}
In embedding coordinates of $AdS_d$, when acting on a totally symmetric rank-$s$ tensor ${\cal U}_{a_1\cdots a_s}$ in the $AdS_d$ direction, the order $R^{-2}=\mu^2$ part of the Laplacian, divergence, and trace operators in \eqref{eqn:AdSgeoCorrection} are given by 
\ie\label{eqn:AdSOpsecondorderPart}
{}&((\delta\nabla^2){\cal U})_{a_1\cdots a_s}=X^bX^c\partial_b\partial_c {\cal U}_{a_1\cdots a_s}+d X^b\partial_b {\cal U}_{a_1\cdots a_s}+s{\cal U}_{a_1\cdots a_s}+2\sum_{j=1}^s X^b\partial_{a_j}{\cal U}_{a_1.\cdots a_{j-1}ba_{j+1}\cdots a_s},
\\
&((\delta\nabla\cdot){\cal U})_{a_2\cdots a_s}=X^bX^c\partial_c {\cal U}_{ba_2\cdots a_s}+(d+s-1)X^b {\cal U}_{ba_2\cdots a_s},
\\
&((\delta\tr){\cal U})_{a_3\cdots a_s}=X^bX^c {\cal U}_{bca_3\cdots a_s}.
\fe

\section{Diffeomorphism analysis} \label{app:Diffeo}
In this appendix, we identify the field redefinition $F_H(T^{(0)})$ in \eqref{eqn:VClinFluc} from the perspective of diffeomorphism in SFT \cite{Mazel:2025fxj}. Rather than trying to investigate the most general diffeomorphism problem at mass-level $n$, which is supposedly a very large problem, our goal is to come up with a reasonable form of $F_H(T^{(0)})$ that is motivated from the covariance of the tensor $T$. Its purpose is merely to provide a useful way to transform the linearized EOM \eqref{eqn:2ndLinEOM} into a set of covariant equations \eqref{eqn:T2equation}.

In SFT, a gauge transformation of a string field $\Psi$ is generated by a ghost-number 1 string field $\Lambda$ is given by
\ie
\delta_{\Lambda}\Psi=Q\Lambda+[e^{\otimes\Psi}\otimes\Lambda].
\fe
For a mass-level $n$ string field $\Phi_n$ that is a linearized fluctuation around $\Psi$, we write $\Lambda=\lambda_0+\lambda_n$ where $\lambda_0$ and $\lambda_n$ are generators of mass-level zero and $n$ respectively. In particular,
\ie
  \lambda_0 = V_{\mu}(X)c\bar{c}\left(e^{-\phi}\psi^{\mu}e^{-2\bar{\phi}}\bar{\partial}\bar{\xi}+e^{-2\phi}\partial\xi e^{-\bar{\phi}}\bar{\psi}^{\mu}\right),
\fe
generates the usual diffeomorphism with generator ${1\over8}V_\mu(X)$ \cite{Mazel:2025fxj} that for example leads to the transformation of the field $\mu^2H_{\mu\nu}$ multiplying $V^{\mu\nu}_{NSNS}$ given by
\ie
\delta_V(\mu^2H_{\mu\nu})={1\over8}(\partial_\mu V_\nu+\partial_\nu V_\mu).
\fe
We extend the transformation generated by $V_\mu(X)$ to mass-level $n$ by assuming that $\lambda_n$ is a function of $V_\mu(X)$ just like $\lambda_0$. To the second order in $\mu$, we assume that $\lambda_n$ is linear in $V_\mu$. In terms of $\psi$ and $\varphi_n$, we have
\ie\label{eqn:diffeoleveln}
\delta_{\lambda}\varphi_n=Q\lambda_n+\left[e^{\otimes\psi}\otimes\lambda_0\otimes\varphi_n\right]'_n=Q\lambda_n+\left[\lambda_0\otimes\varphi_n\right]_n'+\cdots.
\fe

We consider the transformation property of $e^{A(\Box-4n)}{\cal V}_n[{\cal U}]$ for totally symmetric rank-$s$ tensor ${\cal U}$ where we introduced a rescaling as in \eqref{eqn:rescaling2nd}. Further replacing $V_\mu\rightarrow e^{-b\Box/2}V_\mu,H_{\mu\nu}\rightarrow e^{-b\Box/2}H_{\mu\nu},\lambda_0\rightarrow e^{-b\Box/2}\lambda_0,\lambda_n\rightarrow e^{A(\Box-4n)}\lambda_n$, \eqref{eqn:diffeoleveln} becomes
\ie\label{eqn:diffLeveln}
\delta_\lambda{\cal V}_n[{\cal U}]=Q\lambda_n+e^{-A(\Box-4n)}\left[e^{-b\Box/2}\lambda_0\otimes e^{A(\Box-4n)}{\cal V}_n[{\cal U}]\right]_n'+\cdots.
\fe
\eqref{eqn:diffLeveln} restricted to the totally symmetric rank-$s$ tensor component leads to a transformation of ${\cal U}$ that differs from the Lie derivative ${\cal L}_{V/8}{\cal U}$. We posit a covariant totally symmetric rank-$s$ tensor $T=T^{(0)}+\mu T^{(1)}+\mu^2T^{(2)}+{\cal O}(\mu^3)$ related to $\cal U$ via a metric-dependent field redefinition as
\ie
{\cal U}=T^{(0)}+\mu T^{(1)}+\mu^2 T^{(2)}+\mu^2 F_H(T^{(0)})+{\cal O}(\mu^3),
\fe
where $F_H(T^{(0)})$ is linear in $H_{\mu\nu}$. We then demand
\ie\label{eqn:diffeoDemand}
\delta_\lambda{\cal V}_n[{\cal U}]&=Q\lambda_n(V,T^{(0)})+e^{-A(\Box-4n)}\left[ e^{-b\Box/2}\lambda_0\otimes{\cal V}_n[T^{(0)}] \right]_n'+{\cal O}(\mu^3)
\\
&={\cal V}_n[{\cal L}_{V/8}T^{(0)}]+\mu^2{\cal V}_n[F_{\delta_V H}(T^{(0)})]+\mu^2\WT{\cal C}_n(\delta_V H,T^{(0)})+{\cal O}(\mu^3),
\fe
where $\WT{\cal C}_n(H,T^{(0)})$ stands for mass-level $n$ string field components other than ${\cal V}_n$ that are linear in $H_{\mu\nu}$, and $V_\mu$ is assumed to be of order $\mu^2$. \eqref{eqn:diffeoDemand} relates the SFT gauge transformation with the expected transformation of the covariant tensor $T$. It provides constraints on $\lambda_n(V,T^{(0)}),F_H(T^{(0)}),$ and $\WT{\cal C}_n(H,T^{(0)})$. 

Since $H_{\mu\nu}(X)$ for $AdS$ backgrounds of interest is quadratic in $X$, we demand \eqref{eqn:diffeoDemand} for $V_\mu(X)$ that is a general cubic polynomial in $X$ and consider $F_H(T^{(0)})$ that has at most two derivatives. Instead of solving for all ingredients of \eqref{eqn:diffeoDemand}, we focus just on the ${\cal V}_n$ component, which we denote by $(\cdots)|_{{\cal V}_n}$. Note that \eqref{eqn:diffeoDemand} does not completely determine $F_H(T^{(0)})$. In particular, any addition of a term linear in linearized Ricci tensor will not affect $F_{\delta_VH}(T^{(0)})$. It is reasonable to expect that there are choices of $\WT{\cal C}_n(H,T^{(0)})$ such that they agree with ${\cal C}_n(T^{(0)})$ in \eqref{eqn:Cngeneralform} up to $Q$-exact terms, but we do not investigate this further. For simplicity, we also assume a particular form of $T^{(0)}$ that is given by $T^{(0)}=e^{ik\cdot X}u^{\otimes s}$ obeying $k^2+4n=k\cdot u=u\cdot u=0$ since that is the form of the tensor we study for LRT spectral problem as in \eqref{eqn:LRT0thsol}.

The string bracket in the first line of \eqref{eqn:diffeoDemand} evaluates to
\ie
e^{-A(\Box-4n)}&\left[ e^{-b\Box/2}\lambda_0\otimes{\cal V}_n[T^{(0)}] \right]_n'\bigg|_{{\cal V}_n}={\cal V}_n\bigg[e^{ik\cdot X}\bigg({1\over8}(\partial^\rho+ik^\rho)R^0_{\mu_1\cdots\mu_s}
\\
&+{n+1\over16}(u^\rho D_{(\mu_1}-2\delta^\rho_{(\mu_1}(u\cdot\partial))R^1_{\mu_2\cdots\mu_s)}+{n\over32}u^\rho(u\cdot\partial)\eta_{(\mu_1\mu_2}R^2_{\mu_3\cdots\mu_s)}\bigg)V_\rho\bigg],
\fe
where
\ie
D_\mu=&\partial_\mu-ik_\mu,
\\
R^0_{\mu_1\cdots\mu_s}=&u_{\mu_1}\cdots u_{\mu_s}+{n\over2}u_{(\mu_1}\cdots u_{\mu_{s-1}}D_{\mu_s)}(u\cdot\partial)+{n(3n-1)\over32}u_{(\mu_1}\cdots u_{\mu_{s-2}}D_{\mu_{s-1}}D_{\mu_s)}(u\cdot \partial)^2
\\
&+{n(n-1)(5n-1)\over576}u_{(\mu_1}\cdots u_{\mu_{s-3}}D_{\mu_{s-2}}D_{\mu_{s-1}}D_{\mu_{s})}(u\cdot\partial)^3,
\\
R^1_{\mu_1\cdots\mu_{s-1}}=&u_{\mu_1}\cdots u_{\mu_{s-1}}+{3n\over8}u_{(\mu_1}\cdots u_{\mu_{s-2}}D_{\mu_{s-1})}(u\cdot\partial)+{n(5n-2)\over96}u_{(\mu_1}\cdots u_{\mu_{s-3}}D_{\mu_{s-2}}D_{\mu_{s-1})}(u\cdot \partial)^2
\\
&+{5n(n-1)(7n-2)\over9216}u_{(\mu_1}\cdots u_{\mu_{s-4}}D_{\mu_{s-3}}D_{\mu_{s-2}}D_{\mu_{s-1})}(u\cdot\partial)^3,
\\
R^2_{\mu_1\cdots\mu_{s-2}}=&u_{\mu_1}\cdots u_{\mu_{s-2}}+{3n-1\over8}u_{(\mu_1}\cdots u_{\mu_{s-3}}D_{\mu_{s-2})}(u\cdot\partial)
\\
&+{(n-1)(5n-1)\over96}u_{(\mu_1}\cdots u_{\mu_{s-4}}D_{\mu_{s-3}}D_{\mu_{s-2})}(u\cdot \partial)^2.
\fe
It is then straightforward to show that a solution to ${\cal V}_n$ component of \eqref{eqn:diffeoDemand} is given by
\ie
{}&(F_H(T^{(0)}))_{\mu_1\cdots\mu_s}=e^{ik\cdot X}\bigg[{1\over2}H_\mu^\mu u_{\mu_1}\cdots u_{\mu_s}-(n+1)u^\rho H_{\rho(\mu_1}u_{\mu_2}\cdots u_{\mu_s)}
\\
&+{in\over2}k^\rho u^\nu(\partial_\nu H_{\rho(\mu_1}+{\partial}_{(\mu_1}H_{\rho\nu}-\partial_\rho H_{\nu(\mu_1})u_{\mu_2}\cdots u_{\mu_s)}-{in\over2}(u\cdot\partial)H_\mu^\mu k_{(\mu_1}u_{\mu_2}\cdots u_{\mu_s)}
\\
&+{3in(n+1)\over4}k_{(\mu_1}(u^\rho(u\cdot\partial) H_{\rho\mu_2}-u^\rho u^\sigma\partial_{\mu_2} H_{\rho\sigma})u_{\mu_3}\cdots u_{\mu_s)}\bigg],
\fe
where symmetrization in indices applies only to $\mu_l$ indices, and $\lambda_n$ such that
\ie
Q\lambda_n\bigg|_{{\cal V}_n}&={\cal V}_n\bigg[e^{ik\cdot X}\bigg( D^+_{(\mu_1}\tilde a u_{\mu_2}\cdots u_{\mu_s)}+D^+_{(\mu_1} a_{\mu_2} u_{\mu_3}\cdots u_{\mu_s)}+D^+_{(\mu_1} a_{\mu_2\mu_3} u_{\mu_4}\cdots u_{\mu_s)}
\\
&+D^+_{(\mu_1} a_{\mu_2\mu_3\mu_4} u_{\mu_5}\cdots u_{\mu_s)}+\tilde r\eta_{(\mu_1\mu_2}u_{\mu_3}\cdots u_{\mu_s)}+\eta_{(\mu_1\mu_2}r_{\mu_3}u_{\mu_4}\cdots u_{\mu_s)}+\eta_{(\mu_1\mu_2}r_{\mu_3\mu_4}u_{\mu_5}\cdots u_{\mu_s)}\bigg)\bigg],
\fe
with
\ie
D^+_\mu&=\partial_\mu+ik_\mu,~~\tilde a = {n+1\over16}(u\cdot V)+{in\over16}(u\cdot\partial)(k\cdot V)-{n\over16}(u\cdot \partial)(\partial\cdot V),
\\
a_\mu&=-{3in(n+1)\over128}k_\mu(u\cdot\partial)(u\cdot V)+{n(3n-1)\over256}k_\mu(u\cdot\partial)^2(k\cdot V)-{3n(n+1)\over128}\partial_\mu(u\cdot\partial)(u\cdot V)
\\
&+{3n(n+1)\over64}(u\cdot\partial)^2V_\mu-{in(3n-1)\over256}k_\mu(u\cdot\partial)^2(\partial\cdot V)+{3in(3n-1)\over256}\partial_\mu(u\cdot\partial)^2(k\cdot V),
\\
a_{\mu\nu}&=-{n(n+1)(5n-2)\over1536}k_\mu k_\nu(u\cdot\partial)^2(u\cdot V)-{in(n-1)(5n-1)\over4608}k_\mu k_\nu(u\cdot\partial)^3(k\cdot V)
\\
&-{in(n+1)(5n-2)\over384}k_{(\mu}\partial_{\nu)}(u\cdot\partial)^2(u\cdot V)+{in(n+1)(5n-2)\over768}k_{(\mu}(u\cdot\partial)^3 V_{\nu)},
\\
a_{\mu\nu\rho}&={5in(n-1)(n+1)(7n-2)\over147456}k_\mu k_\nu k_\rho(u\cdot\partial)^3(u\cdot V),
\\
\tilde r&=-{n\over32}(u\cdot\partial)(u\cdot V),~~~r_\mu={in(3n-1)\over256}k_\mu(u\cdot\partial)^2(u\cdot V)-{n(3n-1)\over256}\partial_\mu(u\cdot\partial)^2(u\cdot V),
\\
r_{\mu\nu}&={n(n-1)(5n-1)\over3072}k_\mu k_\nu(u\cdot\partial)^3(u\cdot V).
\fe
One such $\lambda_n$ is given by
\ie
\lambda_n=N_ne^{ik\cdot X}\left( \tilde a {\cal S}+a_\mu{\cal S}^\mu+a_{\mu\nu}{\cal S}^{\mu\nu}+a_{\mu\nu\rho}{\cal S}^{\mu\nu\rho}+\tilde r{\cal R}+r_\mu{\cal R}^\mu+r_{\mu\nu}{\cal R}^{\mu\nu} \right),
\fe
where
\ie
{}&{\cal S}^{\mu_1\cdots\mu_r}=u_{\nu_{r+1}}\cdots u_{\nu_{2n+1}}{\mathscr P}^{\mu_1\cdots\mu_r\nu_{r+1}\cdots\nu_{2n+1}},~~{\cal R}^{\mu_1\cdots\mu_r}=u_{\nu_{r+1}}\cdots u_{\nu_{2n}}{\mathscr Q}^{\mu_1\cdots\mu_r\nu_{r+1}\cdots\nu_{2n}},
\\
&{\mathscr P}^{\nu_1\cdots\nu_{2n+1}}={1\over2}e^{-\phi}\partial X^{(\nu_1}\cdots\partial X^{\nu_{n-1}}\psi^{\nu_n}\wedge\bar c e^{-\bar\phi}\bar\partial X^{\nu_{n+1}}\cdots\bar\partial X^{\nu_{2n}}\bar\psi^{\nu_{2n+1})},
\\
&{\mathscr Q}^{\nu_1\cdots\nu_{2n}}={1\over2}ce^{-2\phi}\partial\xi\partial\phi\partial X^{(\nu_1}\cdots\partial X^{\nu_{n-1}}\wedge\bar c e^{-\bar\phi}\bar\partial X^{\nu_{n}}\cdots\bar\partial X^{\nu_{2n-1}}\bar\psi^{\nu_{2n})}.
\fe

\section{$Q{\cal V}_n[F_H(T^{(0)})]$ and ${\cal J}$ terms}\label{app:QFH}
In addition to the operators introduced in \eqref{eqn:operatorNotations1}, we further introduce
\ie\label{eqn:moreoperators}
{}&K_n=c\partial^3 ce^{-\phi}(u\cdot\psi)(u\cdot\partial X)^{n-2},~~S_n=c\eta\partial\phi(u\cdot\partial X)^{n-1},~~R_n=c\partial\eta(u\cdot\partial X)^{n-1},
\\
&T_n=ce^{-\phi}\partial\phi(u\cdot\psi)(u\cdot\partial X)^{n-1},~~Q_n=ce^{\phi}\eta\partial\eta(u\cdot\psi)(u\cdot\partial X)^{n-1}.
\fe
Then, $Q{\cal V}_n[F_H(T^{(0)})]$ is given by
\ie\label{eqn:EOMterms1st}
Q{\cal V}_n[F_H(T^{(0)})]=N_ne^{-iEt}\left[c_0^+{\cal P}^{(1)}_{F_H}+{\cal P}^{(2)}_{F_H}\right],
\fe
with
\ie
{}&{\cal P}^{(1)}_{F_H}=(-iEt+4n^2-n+9-2d-2(d-4)(d-5))B_n\bar B_n-{iEz\over2}\left[ B_n\vee\bar B_n^0+nB_n\vee\bar U^0_n\right],
\\
&{\cal P}^{(2)}_{F_H}={\cal A}_n\left[G_n\wedge\bar B_n+nB_n\wedge\bar P_n \right]+{nz^2\over4}\left[ B_n\wedge\bar S_n+B_n\wedge\bar R_n \right]+{n(n-1)z^2\over24}K_n\wedge\bar B_n,
\\
&{\cal A}_n={z\over8(n+1)}[iE(n+1)t-3n^3-9n^2-(d+2)n-d].
\fe
For ${\cal J}$ terms, we have
\ie\label{eqn:EOMterms2nd}
{\cal J}_w[(\delta\nabla^2)T^{(0)}-M_nT^{(0)}]=&N_ne^{-iEt}c_0^+\bigg[ \left(2nt^2+{idEt\over2}-(n+1)+{M_n\over2}\right)B_n\bar B_n
\\
&-iEzB_n\vee\bar B_n^0-inEzB_n\vee\bar U_n^0\bigg],
\\
{\cal J}_d[(\delta\nabla\cdot)T^{(0)}]=&{N_nze^{-iEt}\over4}(2n+d+1-iEt)(nP_n\wedge\bar B_n-G_n\wedge\bar B_n),
\\
{\cal J}_t[(\delta\tr)T^{(0)}]=&N_ne^{-iEt}z^2\left[ {n(n-1)\over24}K_n\wedge\bar B_n-{n\over4}R_n\wedge\bar B_n-{n\over4}S_n\wedge\bar B_n \right].
\fe

\section{RR source term computation}\label{app:RRsource}
In this appendix, we present details of the computation for \eqref{eqn:4ptRR} that determines the RR source term contribution to the second order linearized EOM. Modulo $(1-\mathbb{P})$ projector in $h$, \eqref{eqn:4ptRR} is nothing but the ordinary on-shell four-point amplitude. Flat superstring vertex choice puts one holomorphic and one anti-holomorphic PCO's on $\tilde\varphi_n^{(0)}$, and we further choose one of $\psi^{(1)}$ to be the integrated vertex operator. We therefore compute the four-point correlator
\ie \label{eqn:4ptCorr}
{\cal I}_n(x,\bar x)=\langle{\mathcal X}_0\bar{\mathcal X}_0\tilde\varphi_n^{(0)}(\infty)\psi^{(1)}(0)\varphi_n^{(0)}(1)\tilde\psi^{(1)}(x,\bar x)\rangle,
\fe
where $\tilde\psi^{(1)}$ is $\psi^{(1)}$ with $c\bar c$ removed by the prescription described in (\ref{eqn:3bracket}), that is, for $\psi^{(1)}=c\bar cV$, then $\widetilde{\psi}^{(1)}=b_{-1}\bar b_{-1}\psi^{(1)}=-V$.

Among the terms resulting from picture-raising in ${\mathcal X}_0\bar{\mathcal X}_0\tilde\varphi_n^{(0)}$, only a single term involving $(k\cdot\psi)(v\cdot\psi)(v\cdot\partial X)^n(k\cdot\bar\psi)(v\cdot\bar\psi)(v\cdot\bar\partial X)^n$ survives since other terms either have unmatched $\partial X$ or $\bar\partial X$ terms that vanish due to transversality, or $\eta$, $\bar\eta$ that cannot saturate the ghost numbers. As a result, $bc$-ghost and $X$-correlators can be computed straightforwardly. The fermion correlator can also be straightforwardly obtained from $G_a^{\alpha\beta}(x)=\langle e^{-\phi}(a\cdot\psi)(1) e^{-{1\over2}\phi}S^\alpha(0) e^{-{1\over2}\phi}S^\beta(x)\rangle={i\over2}{a^\mu U_\mu^{\alpha\beta}\over x(1-x)}$ since $(k\cdot\psi)(v\cdot\psi)(\infty)$ is a Lorentz current leading to the Ward identity:
\ie
{}&\langle (k\cdot\psi)(v\cdot\psi)(w)e^{-\phi}(u\cdot\psi)(1)e^{-{1\over2}\phi}S^\alpha(0)  e^{-{1\over2}\phi}S^\beta(x)\rangle
\\
&=-{G_k^{\alpha\beta}(x)\over w-1}+{((k\cdot U)(v\cdot L)-(v\cdot U) (k\cdot L))^\alpha_{~\delta}G_u^{\delta\beta}(x)\over8w}+{\left((k\cdot U)(v\cdot L)-(v\cdot U)(k\cdot L)\right)_{~\delta}^{\beta}G_u^{\alpha\delta}(x)\over 8(w-x)}.
\fe

Collecting all correlators, we get
\ie
{\cal I}_n(x,\bar x)={E^2{\cal Z}\over16}{2-x-\bar x\over|x|^2|1-x|^2},~~~{\cal Z}=\begin{cases}
    \tr\left((F^{(1)})^TU_0F^{(1)}U_0\right)=-16,~~~&AdS_5\times S^5
    \\
    -\tr\left((F^{(1)})^TU_0F^{(1)}L_0\right)=-12,~~~&AdS_4\times \mathbb{CP}^3
    \\
    \tr\left((F^{(1)})^TU_0F^{(1)}U_0\right)=-8,~~~&AdS_3\times S^3\times{\cal M}_4
\end{cases}
\fe
We now integrate ${\cal I}_n(x,\bar x)$ over the moduli space, while subtracting off the logarithmic divergence near $x\sim0$ as dictated by the projection $(1-\mathbb{P})$. This exercise has been carried out in Appendix F.1.2 of \cite{Cho:2025coy}, where the moduli integration region was split into ${\cal M}_1=\bigg\{x:|{1\over x}-{1\over2}|<r_0\bigg\}$ and ${\cal M}_2=\bigg\{x:|{1\over x}-{1\over2}|\geq r_0\bigg\}$. Integration of ${\cal I}_n(x,\bar x)$ over ${\cal M}_1$ gives
\ie
\int_{{\cal M}_1}{d^2x\over-2\pi i}{\cal I}_n(x,\bar x)={n{\cal Z}\over4}\cdot2\log{1+4r_0^2\over4},
\fe
with $d^2x=dx\wedge d\bar x$. For integration over ${\cal M}_2$, we turn to the plumbing parameter $q$ related to $x$ by $x={2q\over q-2r_0}$ where ${\cal M}_2=\bigg\{q:|q|\leq1\bigg\}$ and subtract off the logarithmic divergence $\sim{1\over|q|^2}$ to obtain
\ie
{n{\cal Z}\over4}\int_{|q|\leq1}{d^2q\over-2\pi i}\left[ {2\over|q|^2}{1-|q|^2/(4r_0^2)\over |1-q^2/(4r_0^2)|^2}-{2\over |q|^2} \right]=-{n{\cal Z}\over4}\cdot2\log{1+4r_0^2\over4r_0^2}.
\fe
We arrive at
\ie
{\cal Y}={n{\cal Z}\over2}\log r_0,
\fe
which gives \eqref{eqn:RRsourceCoeff}.

\section{Determination of ${\cal C}_n(T^{(0)})$}\label{app:Cn}
We show the detailed results for the determination of ${\cal C}_n(T^{(0)})$ in section \ref{subsubsec:Cn}. Let us first briefly summarize the strategy for determining the coefficients. The ansatz given by \eqref{eqn:Cngeneralform} has terms linear in the $X$ zero-modes whose coefficients are $a_{\cal O}$ and other terms quadratic in the $X$ zero-modes with coefficients $b_{\cal O}$ and $c_{\cal O}$. Since BRST charge $Q$ does not increase the degree, string fields in \eqref{eqn:linEOMrequired} that are quadratic in $X$ zero-modes provide equations for $b_{\cal O}$ and $c_{\cal O}$ only, which can be solved for up to a single free parameter $\delta_n$. String fields in \eqref{eqn:linEOMrequired} that are linear in $X$ zero-modes then provide equations for $a_{\cal O}$, which are again determined up to $\delta_n$. However, $a_{\cal O}$ and $b_{\cal O}$ coefficients that appear in front of $c_0^+B_n\bar B_n$ in \eqref{eqn:linEOMrequired} are independent of $\delta_n$ and $M_n$ is therefore also independent of $\delta_n$. In fact, $\delta_n$ can be absorbed into a $Q$-exact term.

Below are explicit expressions for operators appearing in \eqref{eqn:Cngeneralform} for $n\geq2$:
\ie
{\mathfrak M}_n^1:~~&{\cal O}^1_1=B_n\vee\bar F_n,~~{\cal O}^1_2=B_n\vee\bar Y_n,~~{\cal O}^1_3=B_n\vee\bar Z_n,~~{\cal O}^1_4=c_0^+B_n\wedge\bar D_n,~~{\cal O}^1_5=c_0^+B_n\wedge\bar L_n,
\\
{\mathfrak M}_n^2:~~&{\cal O}^2_1=B_n^0\vee\bar F_n,~~{\cal O}^2_2=B_n^0\vee\bar Y_n,~~{\cal O}^2_3=B_n^0\vee\bar Z_n,~~{\cal O}^2_4=U_n^0\vee\bar F_n,~~{\cal O}^2_5=U_n^0\vee\bar Y_n,
\\
&{\cal O}^2_6=U_n^0\vee\bar Z_n,~~{\cal O}^2_7=F_n\bar F_n,~~{\cal O}^2_8=F_n\vee\bar Y_n,~~{\cal O}^2_9=F_n\vee\bar Z_n,~~{\cal O}^2_{10}=Y_n\bar Y_n,
\\
&{\cal O}^2_{11}=Y_n\vee\bar Z_n,~~{\cal O}^2_{12}=Z_n\bar Z_n,~~{\cal O}^2_{13}=G_n\wedge\bar D_n,~~{\cal O}^2_{14}=G_n\wedge\bar L_n,~~{\cal O}^2_{15}=P_n\wedge\bar D_n,
\\
&{\cal O}^2_{16}=P_n\wedge\bar L_n,~~{\cal O}^2_{17}=c_0^+B_n^0\wedge\bar D_n,~~{\cal O}^2_{18}=c_0^+B_n^0\wedge\bar L_n,~~{\cal O}^2_{19}=c_0^+U_n^0\wedge\bar D_n,
\\
&{\cal O}^2_{20}=c_0^+U_n^0\wedge\bar L_n,~~{\cal O}^2_{21}=c_0^+F_n\wedge \bar D_n,~~{\cal O}^2_{22}=c_0^+F_n\wedge \bar L_n,~~{\cal O}^2_{23}=c_0^+ Y_n\wedge\bar D_n,
\\
&{\cal O}^2_{24}=c_0^+Y_n\wedge\bar L_n,~~{\cal O}^2_{25}=c_0^+Z_n\wedge\bar D_n,~~{\cal O}^2_{26}=c_0^+Z_n\wedge\bar L_n.
\fe
We write \eqref{eqn:Cngeneralform} as
\ie
{\cal C}_n(T^{(0)})=N_ne^{-iEt}\left[ z\sum_{l=1}^5(a_l+tb_l){\cal O}^1_l+z^2\sum_{l=1}^{26}c_l{\cal O}^2_l \right].
\fe
$Q$ action on ${\cal C}_n(T^{(0)})$ produces
\ie
Q{\cal C}_n(T^{(0)})=N_ne^{-iEt}\left[ c_0^+{\cal P}^{(1)}_{QC}+{\cal P}^{(2)}_{QC} \right],
\fe
with\footnote{See \eqref{eqn:moreoperators} for definitions of $T_n$ and $Q_n$.}
\ie
{}&{\cal P}^{(1)}_{QC}=\left[-a_4+2a_5+t(-b_4+2b_5)\right]
B_n\bar B_n-iE c_{17}z^2 B_n^0\bar B_n^0+2iE c_{20}z^2 U_n^0\bar U_n^0+z^2(-nc_{21}+2c_{22})F_n\bar F_n
\\
&+2(n-1)c_{24}z^2 Y_n\bar Y_n+{z\over 2}
\left[b_4-iE(a_4+tb_4)-2c_{17}+4c_{18}\right]
B_n\vee\bar B_n^0
\\
&+z
\left[iE(a_5+tb_5)-b_5-c_{19}+2c_{20}\right]
B_n\vee\bar U_n^0
+z\left[
-iEb_1-{n\over 2}(a_4+tb_4)
+a_5+tb_5-c_{21}+2c_{22}
\right]
B_n\vee\bar F_n
\\
&+z\left[
-iEb_2+(n-1)(a_5+tb_5)-c_{23}+2c_{24}
\right]
B_n\vee\bar Y_n+z(-iEb_3-c_{25}+2c_{26})
B_n\vee\bar Z_n
\\
&-{z\over 2}
\left[n(a_4+tb_4)+2(a_5+tb_5)\right]
B_n\vee\bar T_n+{iEz^2\over 2}(2c_{18}-c_{19})
B_n^0\vee\bar U_n^0-{z^2\over 2}(iEc_{21}+nc_{17}-2c_{18})
B_n^0\vee\bar F_n
\\
&+{z^2\over 2}\left[-iEc_{23}+2(n-1)c_{18}\right]
B_n^0\vee\bar Y_n
-{iEc_{25}z^2\over 2}
B_n^0\vee\bar Z_n
-{z^2\over 2}(nc_{17}+2c_{18})
B_n^0\vee\bar T_n
\\
&-{z^2\over 2}(-2iEc_{22}+nc_{19}-2c_{20})
U_n^0\vee\bar F_n
+z^2\left[iEc_{24}+(n-1)c_{20}\right]
U_n^0\vee\bar Y_n
+iEc_{26}z^2
U_n^0\vee\bar Z_n
\\
&-{z^2\over 2}(nc_{19}+2c_{20})
U_n^0\vee\bar T_n
+{z^2\over 2}
\left[2(n-1)c_{22}-nc_{23}+2c_{24}\right]
F_n\vee\bar Y_n
-{z^2\over 2}(nc_{25}-2c_{26})
F_n\vee\bar Z_n
\\
&-{z^2\over 2}(nc_{21}+2c_{22})
F_n\vee\bar T_n
+(n-1)c_{26}z^2
Y_n\vee\bar Z_n
-{z^2\over 2}(nc_{23}+2c_{24})
Y_n\vee\bar T_n
\\
&-{z^2\over 2}(nc_{25}+2c_{26})
Z_n\vee\bar T_n
-{iz^2\over 4}(Ec_{17}-2ic_{21})
G_n\wedge\bar D_n
-{iz^2\over 4}(Ec_{18}-2ic_{22})
G_n\wedge\bar L_n
\\
&+{iz^2\over 4}(Ec_{19}-2ic_{21}-4ic_{23})
P_n\wedge\bar D_n
+{iz^2\over 4}(Ec_{20}-2ic_{22}-4ic_{24})
P_n\wedge\bar L_n
-{z^2\over 2}(nc_{17}-c_{19}+c_{25})
A_n\wedge\bar D_n
\\
&-{z^2\over 2}(nc_{18}-c_{20}+c_{26})
A_n\wedge\bar L_n
-{z^2\over 4}
\left[-iEc_{25}+2(n-1)c_{21}-4c_{23}\right]
H_n\wedge\bar D_n
\\
&-{z^2\over 4}
\left[-iEc_{26}+2(n-1)c_{22}-4c_{24}\right]
H_n\wedge\bar L_n,
\fe
and
\ie
{}&{\cal P}_{QC}^{(2)}
=
{z\over 4}
\left[
2a_1-a_4+t(2b_1-b_4)-4c_{13}+8c_{14}
\right]
G_n\wedge\bar B_n
+{z^2\over 4}
\left[-2iEc_{13}+2c_1-c_{17}\right]
G_n\wedge\bar B_n^0
\\
&+{z^2\over 4}
\left[4iEc_{14}-c_{19}+2c_4\right]
G_n\wedge\bar U_n^0
+{iz^2\over 4}
\left[
Ec_1+2inc_{13}-4ic_{14}+ic_{21}-2ic_7
\right]
G_n\wedge\bar F_n
\\
&+{iz^2\over 4}
\left[
Ec_2-4i(n-1)c_{14}+ic_{23}-2ic_8
\right]
G_n\wedge\bar Y_n
+{iz^2\over 4}
\left[Ec_3+ic_{25}-2ic_9\right]
G_n\wedge\bar Z_n
\\
&-{z^2\over 2}(nc_{13}+2c_{14})
G_n\wedge\bar T_n
-{z\over 2}
\left[
a_1+2a_2+a_5+t(b_1+2b_2+b_5)+2c_{15}-4c_{16}
\right]
P_n\wedge\bar B_n
\\
&-{z^2\over 2}
\left[iEc_{15}+c_1+c_{18}+2c_2\right]
P_n\wedge\bar B_n^0
-{z^2\over 2}
\left[-2iEc_{16}+c_{20}+c_4+2c_5\right]
P_n\wedge\bar U_n^0
\\
&-{iz^2\over 4}
\left[
Ec_4-2inc_{15}+4ic_{16}-2ic_{22}-2ic_7-4ic_8
\right]
P_n\wedge\bar F_n
\\
&-{iz^2\over 4}
\left[
Ec_5-4ic_{10}+4i(n-1)c_{16}-2ic_{24}-2ic_8
\right]
P_n\wedge\bar Y_n
-{iz^2\over 4}
\left[Ec_6-4ic_{11}-2ic_{26}-2ic_9\right]
P_n\wedge\bar Z_n
\\
&-{z^2\over 2}(nc_{15}+2c_{16})
P_n\wedge\bar T_n
+{z\over 2}(a_3+tb_3)
A_n\wedge\bar B_n
+{c_3z^2\over 2}
A_n\wedge\bar B_n^0
+{c_6z^2\over 2}
A_n\wedge\bar U_n^0
\\
&+{z^2\over 2}(nc_1-c_4+c_9)
A_n\wedge\bar F_n
+{z^2\over 2}(nc_2-c_5+c_{11})
A_n\wedge\bar Y_n
+{z^2\over 2}(nc_3-c_6+c_{12})
A_n\wedge\bar Z_n
\\
&+{z\over 4}
\left[
-iE(a_3+tb_3)
+2(n-1)(a_1+tb_1)
-4(a_2+tb_2)+b_3
\right]
H_n\wedge\bar B_n
\\
&+{z^2\over 4}
\left[-iEc_3+2(n-1)c_1-4c_2\right]
H_n\wedge\bar B_n^0
+{z^2\over 4}
\left[-iEc_6+2(n-1)c_4-4c_5\right]
H_n\wedge\bar U_n^0
\\
&+{z^2\over 4}
\left[-iEc_9+2(n-1)c_7-4c_8\right]
H_n\wedge\bar F_n
+{z^2\over 4}
\left[-iEc_{11}-4c_{10}+2(n-1)c_8\right]
H_n\wedge\bar Y_n
\\
&+{z^2\over 4}
\left[-iEc_{12}-4c_{11}+2(n-1)c_9\right]
H_n\wedge\bar Z_n
+{z^2\over 2}(nc_{13}+c_{15})
Q_n\wedge\bar D_n
+{z^2\over 2}(nc_{14}+c_{16})
Q_n\wedge\bar L_n.
\fe

With \eqref{eqn:EOMterms1st}, \eqref{eqn:EOMterms2nd}, \eqref{eqn:NSsourceterm}, and \eqref{eqn:RRsourceterm}, we can solve for undetermined $a_l,b_l,c_l$ coefficients from \eqref{eqn:linEOMrequired}. From quadratic terms in $X$ zero-modes, we get an one-parameter family of solutions labeled by $\delta_n$:
\ie
{}&b_1=b_2=b_3=0,~~b_4=-{iE\over2},~~b_5={iEn\over4},
\\
&c_3=c_6=c_9=c_{11}=c_{12}=c_{25}=c_{26}=0,~~c_1={i\delta_n\over E},~~c_2={i\delta_n(n-1)\over2E},~~c_4={iE\delta_n\over4},
\\
&c_5={iE\delta_n(n-1)\over8},~~c_7=\delta_n,~~c_8={\delta_n(n-1)\over2},~~c_{10}={\delta_n(n-1)^2\over4},~~c_{13}={\delta_n\over4n}+{1\over16},
\\
&c_{14}=-{\delta_n\over8}-{n\over32},~~c_{15}=-{\delta_n\over4}-{n\over16},~~c_{16}={\delta_n n\over8}+{n^2\over32},~~c_{17}=-{iE\over8},~~c_{18}={iEn\over16},
\\
&c_{19}=-{iEn\over8},~~c_{20}={iEn^2\over16},~~c_{21}=-{n\over4},~~c_{22}={n^2\over8},~~c_{23}=-{n(n-1)\over8},~~c_{24}={n^2(n-1)\over16}.
\fe
These in turn determine $a_l$'s from terms in \eqref{eqn:linEOMrequired} linear in $X$ zero-modes as
\ie
{}&a_1={\delta_n(n+1)\over2n}+{n(n+2)(3n-1)\over4n+4},~~a_2={\delta_n(n^2-1)\over4n}+{n(3n^3+3n^2-5n+3)\over8n+8},
\\
&a_3={iEn\over4},~~a_4={d+n+3\over4},~~a_5=-{n(d+n+3)\over8}.
\fe
It is straightforward to show that these coefficients indeed solve the full equations \eqref{eqn:linEOMrequired} once $M_n$ is set to \eqref{eqn:massCorrection}. Furthermore, the freedom $\delta_n$ in these solutions corresponds to a $Q$-exact term:
\ie
{}&{\cal C}_n(T^{(0)})={\cal C}_n(T^{(0)})\bigg|_{\delta_n=0}+\delta_nQ\Lambda_n^{C},
\\
&\Lambda_n^{C}=N_ne^{-iEt}z^2\left[ {1\over2n}F_n\wedge\bar D_n-{1\over4}F_n\wedge\bar L_n+{n-1\over4n}Y_n\wedge\bar D_n-{n-1\over8}Y_n\wedge\bar L_n \right].
\fe

For $n=1$, we have
\ie
{\cal C}_1(T^{(0)})={\cal C}_n(T^{(0)})\bigg|_{n=1,Y_1=Z_1=0}+{8e^{-2it}z\over4}\left[ B_1\vee\bar F_1-B_1\vee\bar T_1 \right],
\fe
while for the gauge parameter, we can simply take $n=1$ expression of $\Lambda^C_n$.

Note that the coefficient of the leading Regge component is independent of $\delta_n$:
\ie\label{eqn:CnLRTcomp}
Q{\cal C}_n(T^{(0)})&=N_ne^{-iEt}c_0^+B_n\bar B_n\left(-a_4+2a_5+t(-b_4+2b_5)\right)+\cdots
\\
&=N_ne^{-iEt}c_0^+B_n\bar B_n\left( {iE(n+1)\over2}t-{(n+1)(n+d+3)\over4} \right)+\cdots.
\fe

\bibliographystyle{JHEP}
\bibliography{refs}

\end{document}